\RequirePackage{etoolbox}
\patchcmd{\bibliographystyle}{#1}{naturemagnourlnoarxiv}{}{}

\documentclass[12pt]{wlscirep}
\usepackage[utf8]{inputenc}
\usepackage[T1]{fontenc}

\usepackage{graphicx} 
\usepackage{float}    
\usepackage{subfig}   

\usepackage[mathscr]{euscript} 

\newsavebox{\foobox}
\newcommand{\slantbox}[2][0]{\mbox{%
        \sbox{\foobox}{#2}%
        \hskip\wd\foobox
        \pdfsave
        \pdfsetmatrix{1 0 #1 1}%
        \llap{\usebox{\foobox}}%
        \pdfrestore
}}
\newcommand\unslant[2][-.15]{\slantbox[#1]{$#2$}}
\newcommand{\micron}{$\unslant\mu$m}

\usepackage{fancyhdr}
\newcommand{\SI}[1]{Sec.~\ref{#1}:~\nameref{#1}}

\title{An optic to replace space and its application towards ultra-thin imaging systems}

\author[1,*]{Orad~Reshef}
\author[1]{Michael~P.~DelMastro}
\author[1]{Katherine~K.~M.~Bearne}
\author[1]{Ali~H. Alhulaymi}
\author[1,2]{Lambert~Giner}
\author[1,3,4]{Robert~W.~Boyd}
\author[1]{Jeff~S.~Lundeen}
\affil[1]{Department of Physics, University of Ottawa, 25 Templeton Street, Ottawa, ON \ K1N 6N5, Canada}
\affil[2]{D\'{e}partement de Physique et d'Astronomie, Universit\'{e} de Moncton, Moncton, New Brunswick E1A 3E9, Canada}
\affil[3]{School of Electrical Engineering and Computer Science, University of Ottawa, Ottawa, ON \ K1N 6N5, Canada}
\affil[4]{Institute of Optics and Department of Physics and Astronomy, University of Rochester, Rochester, New York 14627, USA}
\affil[*]{e-mail: orad@reshef.ca}

\begin{abstract}

Centuries of effort to improve imaging has focused on perfecting and combining lenses to obtain better optical performance and new functionalities. The arrival of nanotechnology has brought to this effort engineered surfaces called metalenses, which promise to make imaging devices more compact. However, unaddressed by this promise is the space between the lenses, which is crucial for image formation but takes up by far the most room in imaging systems. Here, we address this issue by presenting the concept of and experimentally demonstrating an optical `spaceplate', an optic that effectively propagates light for a distance that can be considerably longer than the plate thickness.  Such an optic would  shrink future imaging systems, opening the possibility for ultra-thin monolithic cameras. More broadly, a spaceplate can be applied to miniaturize important devices that implicitly manipulate the spatial profile of light, for example, solar concentrators, collimators for light sources, integrated optical components, and spectrometers.

\end{abstract}

\begin{document}

\flushbottom
\maketitle

\subsection*{Introduction}
Metasurfaces --- engineered surfaces consisting of sub-wavelength scatterers --- have attracted a great deal of attention for enabling flat optical components~\cite{Yu2011,Kildishev2013,Yu2014,Meinzer2014,Chen2016c,Genevet2017}. These devices have been implemented in a diverse set of novel linear~\cite{Cui2014,Karimi2014,Stillinger2015,Arbabi2017a,Faraji-Dana2018} and nonlinear optical~\cite{Yang2015,Li2015c,Li2017d} applications, including sub-wavelength-scale broadband achromatic lenses~\cite{Chen2018}, the generation of various transverse spatial modes~\cite{Yu2011,Karimi2014}, lasing~\cite{Zhou2013,Xu2017}, polarimetry~\cite{Rubin2019}, and holograms~\cite{Ni2013}, among others. Notably, metalenses are seen as the most promising by far due to their impact in miniaturizing imaging systems~\cite{Khorasaninejad2017,Banerji2019}. However, in all imaging systems, lenses represent just one of the two main components; the other, sometimes overlooked in this context, is the millimeter-to-meter-scale optical propagation surrounding the lenses and separating them from the object and image. As evidenced by the long physical length of a typical (\emph{e.g.,} Galilean) telescope, the distances between lenses are just as critical to image formation as the lenses themselves, and can easily be greater than the summed thicknesses of the lenses by an order of magnitude. To date, no work has been published that addresses this dominant contribution to the size of many optical systems. 

\begin{figure}[htbp]
\centering \includegraphics[width=0.75\textwidth]{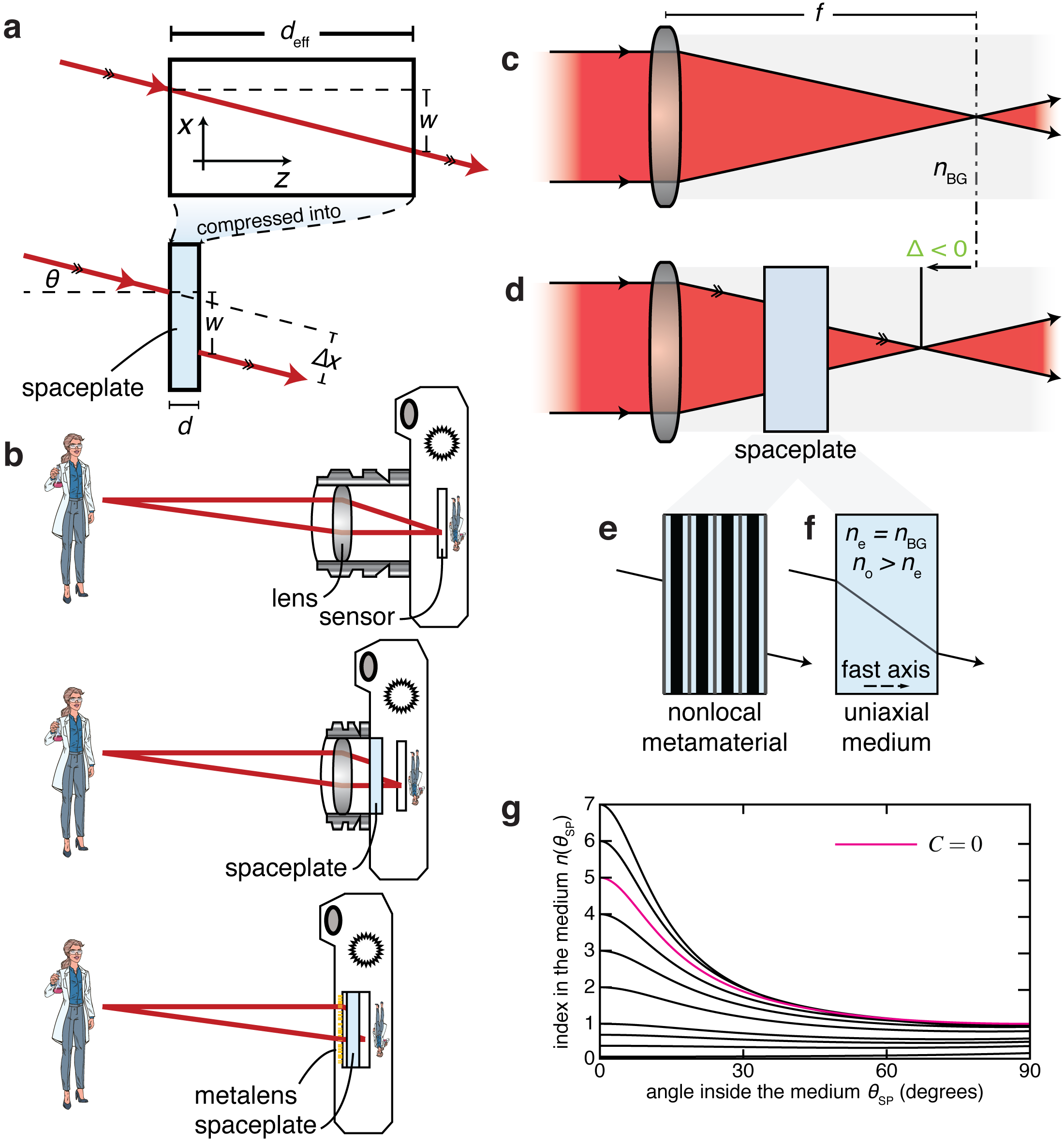}
\caption{\textbf{~|~Operating principle of a spaceplate.} \textbf{a,~}A spaceplate can compress a propagation length of $d_{\mathrm{eff}}$ into a thickness $d$. For example, a beam incident on the spaceplate at angle~$\theta$ will emerge at that same angle and be transversely translated by length~$w$ (resulting in a lateral beam shift~$\Delta x$), just as it would for $d_{\mathrm{eff}}$ of free space. \textbf{b,~}Adding a spaceplate to an imaging system such as a standard camera (top) will shorten the camera (center). An ultrathin monolithic imaging system can be formed by integrating a metalens and a spaceplate directly on a sensor (bottom).  \textbf{c,~}A lens focuses a collimated beam at a working distance corresponding to its focal length~$f$. \textbf{d,~} A spaceplate will act to shorten the distance from the lens to the focus by a distance $|\Delta|$.  The emerging rays are parallel to the original incident rays, which preserves the lens strength. The plate therefore effectively propagates light for a longer length than the physical space it occupies. This effect can be achieved using \textbf{e,} a nonlocal metamaterial, or \textbf{f,} for the extraordinary ray for propagation along the fast axis (e) of a uniaxial birefringent medium with $n_{\mathrm{BG}}=n_{\mathrm{e}}$. \textbf{g,~}A spaceplate can be made of a homogeneous medium with any of these angle-dependent refractive index curves, parametrized by the quantity $C$.}
\label{Fig:schematic}
\end{figure}

We present here a potential path towards replacing these distances with an optical element that we call a `spaceplate'. The functionality of a spaceplate is highlighted in Fig.~\ref{Fig:schematic}a--- this element would occupy a physical thickness of $d$ while propagating light for an effective length of $d_{\mathrm{eff}}>d$, where the ratio between these two quantities $\mathscr{R}\equiv{d_{\mathrm{eff}}/d}$ is the compression factor of the plate. Some metamaterials, such as those based on transformation  optics~\cite{Leonhardt2006,Pendry2006}, already feature the compression of electromagnetic fields (\emph{e.g.,} for field concentrators~\cite{Rahm2008} or hyperlenses~\cite{Kildishev2007}). Though compression of propagation distance can be implicit in these works, this compression has not been an aim unto itself, particularly for reducing the length of imaging systems. There are many optical devices that implicitly use imaging (\emph{e.g.,} a grating spectrometer works by imaging a slit~\cite{czerny1930astigmatismus}) or that spatially manipulate light using its propagation, such as solar concentrators~\cite{Winston1974}, multiplane mode demultiplexers~\cite{Armstrong2012}, or multi-mode interferometers in integrated optics~\cite{Halir2016}. All of these devices could be shortened through use of a spaceplate, leading to significant practical advantages.

It is easiest to describe the operation of the spaceplate by way of example --- to this end, we consider the use of a spaceplate in a camera, as illustrated in Fig.~\ref{Fig:schematic}b. The space between the lens and the sensor of a camera is dictated to a large degree by the focal length $f$ of the lens. A relatively large focal length is necessary to suitably magnify an image, which leads to long lens-barrels in cameras. One approach towards reducing this length could be the use of a spaceplate, allowing for the large magnification of a faraway object without the need for a large propagation length. While nominally this is also the goal of a telephoto lens, in practice the length of a telephoto lens barrel has been approximately constrained to be 0.8 times its effective focal length~\cite{Kingslake2010} (See \SI{SEC:Telephoto} for more discussion). Unlike a telephoto lens, a spaceplate could therefore break the trade-off between lens-barrel length and image magnification. Moreover, since the resulting image may now be large, so can the image sensor (\emph{e.g.,} the charge-coupled device (CCD) array). One can capitalize on this larger sensor by using larger pixels for low-light sensitivity, or a greater number of pixels for a higher resolution. In this way, the spaceplate could one day break the trade-off between camera miniaturization and any of resolution, sensitivity, magnification, or field-of-view.

In this work, we develop the general concept and theory of the spaceplate, followed by the proposal of several physically realizable types of spaceplates. In particular, we introduce and fully simulate the operation of a proof-of-principle multilayer design (\emph{i.e.,} a `thin-film stack'). This 25-layer design consists of only two materials, barely utilizing the full potential of current fabrication capability, which reaches over a thousand layers of multiple different materials~\cite{Martin2010}. We follow this modelling by some experimental demonstrations of two other types of spaceplates, albeit ones with a modest compression factor. These experiments will establish that spaceplates can be polarization independent, exhibit broadband operation, or have a large numerical aperture; thus, in concept, spaceplates can satisfy all three of these performance targets. However, it remains to be shown that these targets are compatible with each other or with a usefully large compression factor and effective length $d_{\mathrm{eff}}$. We conclude with some discussion and analysis of the further advances that are necessary for spaceplates to become practical devices.

\subsection*{Results}
\subsubsection*{Fourier optics analysis of the operation of a spaceplate}
To define the action of the spaceplate, we make use of Fourier optics~\cite{Goodman2005}. Namely, we consider how free propagation transforms each plane-wave component of an incident field. Each plane-wave is a given transverse spatial Fourier component with momentum vector $\mathbf{k}$. The amplitude of each $k$-vector component is preserved in its free-space propagation whereas its phase is shifted. Consider two points along $z$ separated by $d_{\mathrm{eff}}$ for a given plane-wave. The wave's phase difference between these points will be $\phi=k_{z}d_{\mathrm{eff}}$, where $k_{z}=|\mathbf{k}|\cos\theta$, and $\theta$ is the angle of $\mathbf{k}$ from the $z$-axis. Combining this amplitude and phase behaviour, the Fourier transfer function of free space is $H(\mathbf{k})=\exp({ik_{z}d_{\mathrm{eff}}})$. Free propagation will effectively multiply each incident plane-wave by this factor.

A spaceplate needs to produce the same transfer function. A transfer function $H(\mathbf{k})$ with the $k$-vector $|\mathbf{k}|=(2\pi n_{\mathrm{BG}}/\lambda)$ yields a propagation phase of $\phi=(2\pi n_{\mathrm{BG}}d_{\mathrm{eff}}\cos\theta/\lambda)\equiv \phi_\mathrm{BG}$, where $\lambda$ is the wavelength of light in vacuum and $n_{\mathrm{BG}}$ is the index of the background medium (BG) in the $d_\mathrm{eff}$ slab of space. The critical action of a spaceplate is thus to produce an angle-dependent phase profile $\phi_\mathrm{SP}$ that is equal to $\phi_\mathrm{BG}(\theta,d_{\mathrm{eff}})$, the phase from propagation through a distance $d_\mathrm{eff}$ of the background medium.  However, the spaceplate must do so within a distance shorter than $d_{\mathrm{eff}}$ --- in particular, in a plate thickness $d$. Note that the angular phase profile $\phi_\mathrm{SP}(\theta)$ possesses the following two properties. The first is that the addition of an arbitrary phase offset $\phi_{\mathrm{G}}$ that is global (\emph{i.e.,} independent of $\theta$) will not affect the imaging properties of the system~\cite{Banerji2019}. Second, the image will also not be affected if $\phi_\mathrm{SP}(\theta)$ is discontinuous as a function of $\theta$ with discontinuities of an integer multiple $m$ of $2\pi$; this type of solution would correspond to the Fourier-space analogue of a Fresnel lens~\cite{Fresnel,Miyamoto1961}. These two free parameters hint at the substantial flexibility available to design a spaceplate. 

Such a momentum-dependent response, where an optical element acts on the phase or magnitude of the spatial Fourier components of a beam, has been called a `nonlocal' response~\cite{Castaldi2012a,Silva2014,Kwon2018}. Specifically, an ideal spaceplate would impart the phase, 
\begin{equation}
    \phi_{\mathrm{SP}}(k_{x},k_{y},d_{\mathrm{eff}})=d_{\mathrm{eff}}(|\mathbf{k}|^{2}-k_{x}^{2}-k_{y}^{2})^{1/2},\label{EQ:Spaceplate}
\end{equation}
whereas its `local' position-dependent counterpart is a positive, spherical, thin lens, $\phi_{\mathrm{lens}}(x,y,f)=(2\pi/\lambda)(f^{2}-x^{2}-y^{2})^{1/2}$~\cite{Khorasaninejad2016}.  Unlike a  position-dependent response, a purely momentum-dependent response, such as the one in Eq.~\ref{EQ:Spaceplate}, cannot redistribute momentum components. That is, it cannot redirect the angle of a light-ray and, thus, it comes with no magnification and has no optical power (\emph{i.e.,} dioptric power), unlike curved mirrors or lenses.  Therefore, a spaceplate is an optical element complementary to the lens.

Nonlocal response engineering has been a fruitful research direction baring applications such as angular pass-filtering~\cite{Shen2014}, image processing~\cite{Zhou, Zhang2019a, Dong2018, Chazot2020}, and analog computing~\cite{Silva2014}. One previous work in nonlocal responses used a lens system similar to a 4f telescope to impart a  $k$-dependent phase \textit{and} magnitude response~\cite{Silva2014}. However, the use of a lens and propagation to create a spaceplate defeats its purpose of replacing propagation. 
A metamaterial, on the other hand, has only ever been engineered to have an angle-dependent transmittance~\cite{Silva2014, Shen2014,Zhou,Zhang2019a,Dong2018}, thereby solely affecting the \textit{magnitude} of the Fourier component. In contrast, we focus on materials that impart a \textit{phase} to each Fourier component. In order to achieve this behaviour, we consider spaceplate designs that are translationally invariant along the transverse directions $x$ and $y$. This invariance guarantees that a transmitted wave will have the same $k$-vector as the incident wave, which is a necessity for unity transmittance, $|H|=1$. By manipulating the momentum-dependent phase, the spaceplate is a first example along a new avenue in nonlocal metamaterials research.

\subsubsection*{A multilayer spaceplate design}
Since nonlocal responses are based in momentum-space, and not in position-space such as with metasurfaces, it is at first glance not obvious whether a nonlocal response corresponding to a spaceplate may be realized in a physical system, or whether a realistic spaceplate would have any intrinsic trade-offs between its performance parameters. We now explore whether a spaceplate can be designed out of multilayer stack (Fig.~\ref{Fig:schematic}e). Since this structure is made up of parallel flat layers of various materials, it possesses the transverse translational-invariance that we desire. Moreover, the production of these stacks is a mature technology, appearing in many consumer and industrial products, with commercial companies capable of fabricating sophisticated designs with thousands of layers of several different materials. Consequently, such stacks can incorporate considerable complexity and design freedom. In Ref.~\cite{Silva2014},  multilayer stacks were theoretically considered for general nonlocal responses, and a structure was designed that, in modelling, modulated the momentum-dependent transmittance magnitude. Instead, we design a stack to impart the momentum-dependent phase $\phi_\mathrm{BG}(\theta,d_{\mathrm{eff}})$ that we require for a spaceplate (where $d_{\mathrm{eff}}$ is greater than $d$,  the total stack thickness). The purpose of this design is simply to establish that a multilayer architecture can produce a spaceplate and also to determine some initial performance characteristics.

\begin{figure}[b!]
\centering \includegraphics[width=\textwidth]{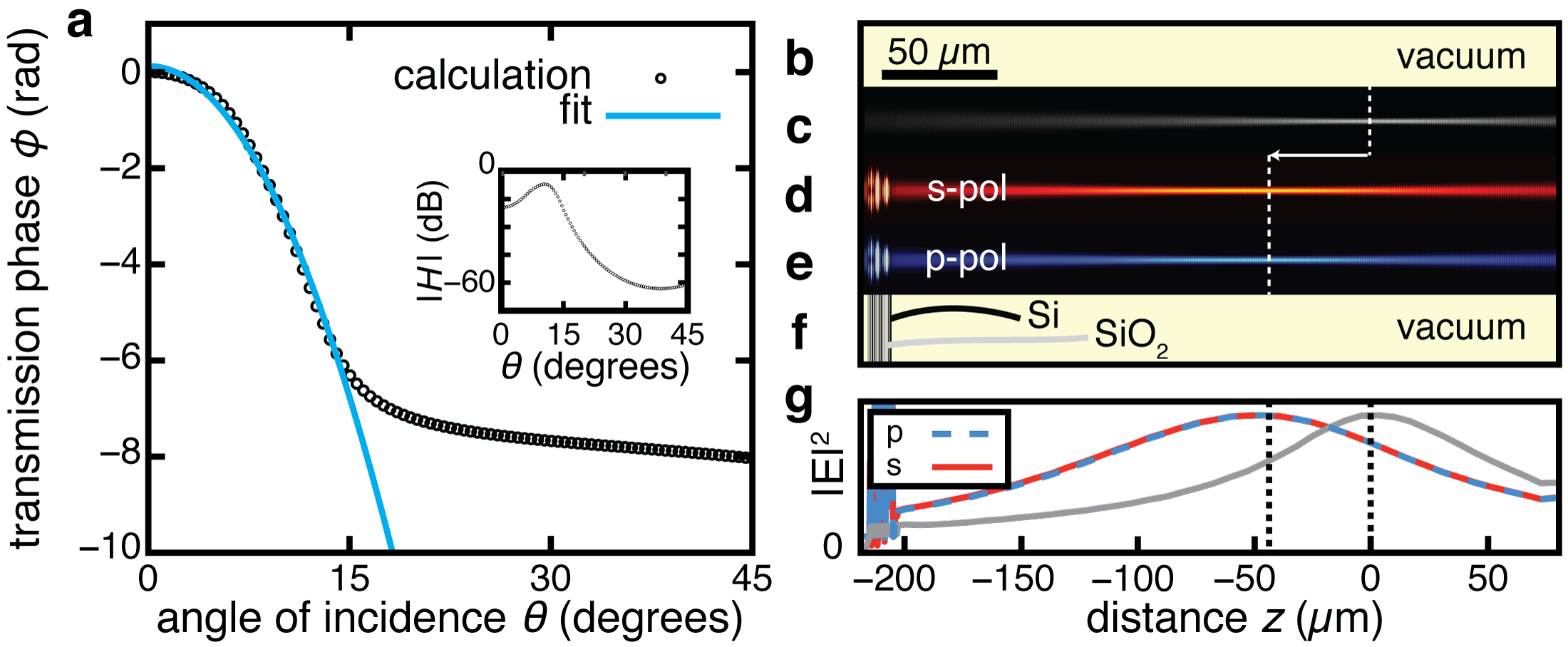}
\caption{\textbf{~|~A nonlocal metamaterial spaceplate. a,~}A multilayer stack consisting of alternating layers of silicon and silica of various thicknesses is engineered to reproduce the Fourier transfer function $H$ for propagation through vacuum for incident angles smaller than $\theta=15^{\circ}$ at an optical wavelength of $\lambda=1550$~nm. Plotted is the calculated transmission phase $\phi_\mathrm{SP}$ of the metamaterial spaceplate (black circles) and a fitted vacuum transfer function phase $\phi_\mathrm{BG}$ (blue curve). Here, we have subtracted a global phase of $\phi_{G}=-0.05$~rad. The fitted compression factor is $\mathscr{R}=4.9$. The inset shows the transmission amplitude $|H|$. \textbf{(c~--e)~}Full-wave simulations of the square of the magnitude of the electric field, $|E|^2$, of a focusing Gaussian beam (waist of $3\lambda$, divergence of $6^{\circ}$) propagating in  \textbf{c,~}vacuum (grey),  \textbf{d,~}after propagating an s-polarized beam through the metamaterial (red, to scale), and \textbf{e,~}after propagating a p-polarized beam through the metamaterial (blue, to scale).  \textbf{(b, f)~} The physical layouts of the simulations, to scale. \emph{i.e.,} \textbf{b,~}is vacuum and \textbf{f,~}is the spaceplate structure surrounded by vacuum. \textbf{g,~}A cross section of $|E|^{2}$ along the beam axis. Transmission through the spaceplate advances the focus position along $z$ by $\Delta=-43.2$~\micron{} for both p-polarized (dashed blue) and s-polarized (solid red) light.}
\label{Fig:metamaterial}
\end{figure}

Since a deterministic and analytic design method for general multilayer stacks has yet to be invented, we use an optimization-based design method, as in Ref.~\citenum{Silva2014}. In particular, we use a genetic algorithm targeting $\phi_\mathrm{BG}(\theta)$ that maximizes the compression factor~$\mathscr{R}$. To set a realistic but relevant goal, we aim only to produce this phase response for a numerical aperture that matches that of modern smartphone cameras, that is, out to an incident angle of $\theta=15^{\circ}$, \emph{i.e.,} NA = 0.26 (Fig.~\ref{Fig:metamaterial}a). Similarly, in order to aim for an easily fabricated design, we restrict the algorithm to two common materials, silica and silicon. This restriction is in contrast to the work in Ref.~\citenum{Silva2014}, where they employed permittivities of arbitrary values idealized to be lossless. We limited our structure to a total thickness of approximately 10~\micron{} and a maximum of 40 layers so that the algorithm could comfortably run on a standard personal computer. The algorithm took four hours to yield a $d\sim 10$~\micron-thick, 25-layer structure.  It acts as a spaceplate with a compression factor of $\mathscr{R}=4.9$ for vacuum-filled space ($n_\mathrm{BG}=1$) for 1550~nm wavelength light. (See \SI{SEC:metamaterial} for more details on this structure.) 
 
In order to test the performance of the designed multilayer structure, we use full-wave simulations to propagate a converging beam through the stack. Full-wave simulations have been validated in the literature to provide accurate predictions for linear optical responses in dielectric materials, such as the case here~\cite{Taflove2005}. Consequently, we do not fabricate and experimentally test this structure, which, with its moderate value of $\mathscr{R}$, small $d_\mathrm{eff}$, and low transmittance, is still far from being a useful device. Simulations of a focusing beam propagating in vacuum and in a spaceplate are shown in Fig.~\ref{Fig:metamaterial}b~--~f.  Analogous to in Fig.~\ref{Fig:schematic}c~--~d, these show that this structure indeed advances the beam focus in vacuum towards the plate, as desired. Fig.~\ref{Fig:metamaterial}g shows the advance is $\Delta=-43.2$~\micron, which corresponds to a compression factor of $\mathscr{R}=5.2$, in approximate agreement with the prediction. Thus, 10~\micron{} of spaceplate metamaterial may be used to replace propagation through over 50~\micron{} of the background medium, here, vacuum.  Though these properties were not explicitly requested by our optimization algorithm, our simulations show that this device design is both polarization-insensitive and advances the focus for a bandwidth spanning 30~nm. (See \SI{SEC:metamaterial} for more details on the performance of this structure.) Crucially, the compression factor $\mathscr{R}$ of this structure exceeds the ratio of any of the indices in the spaceplate ($n_\mathrm{Si}\approx 3.48$, $n_{\mathrm{SiO}_2}\approx 1.45$, $n_\mathrm{vac}=1$) and, thereby, demonstrates that this ratio does not impose a fundamental limit on $\mathscr{R}$. While here we aimed to replace vacuum, in order to achieve a higher numerical aperture for an imaging system, one could instead design the multilayer structure to replace a higher-index background medium since $\mathrm{NA}=n_\mathrm{BG}\sin(\theta)$. The success of the relatively simple structure we designed hints at the promise of more complicated multilayer stacks for creating spaceplates with large compression factors.
 
\begin{figure}[htb]
\centering \includegraphics[width=\textwidth]{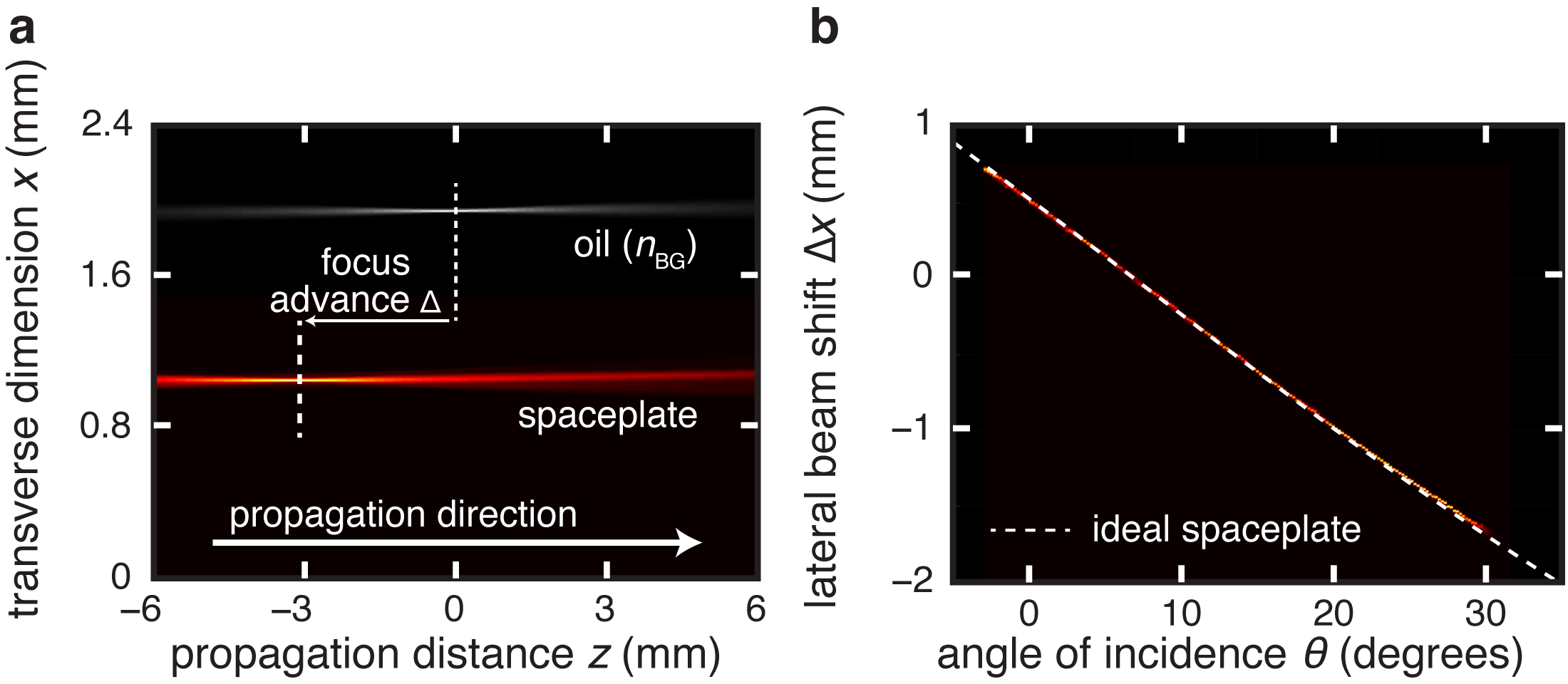}
\caption{\textbf{~|~Experimental demonstration of space compression.} 
For all plots, the false-colour along the plot-vertical gives the transverse intensity distribution along $x$ at each $z$ distance on the horizontal plot axis, with paler colour corresponding to higher intensity.
\textbf{a,~}Focal shift, $\Delta=d-d_{\mathrm{eff}}$. Top data: Oil (grey). A converging beam comes to focus in oil at $z=0$. Bottom data: Uniaxial spaceplate (red). Propagation of an e-polarized beam through a calcite crystal with its fast axis along $z$ advances the focus position by $\Delta=-3.4$~mm. The corresponding $y$ intensity distributions are shown in Sec.~\ref{SEC:2D_spaceplate} in the Supplementary Information, demonstrating a fully two-dimensional advance.  \textbf{b,~}The walk-off of a beam incident at an angle $\theta$. The  dashed line give the lateral beam shift for an ideal spaceplate (\emph{i.e.,} $\Delta x=-(\mathscr{R}-1)d\sin\theta$ ) with the same thickness $d$ and compression factor $\mathscr{R}$ as the spaceplate in \textbf{(a)}. The uniaxial birefringent crystal acts as a perfect spaceplate for all measured angles of incidence.}
\label{Fig:beams_measurement}
\end{figure}
 
\subsubsection*{Homogeneous media as a spaceplate}
A drawback of multilayer stacks is the need for an optimization-based design method, which provides little physical insight into the limitations and operating mechanisms of a spaceplate. For this reason, in this section, we instead consider the possibility of unstructured spaceplates, \emph{i.e.,} homogeneous media. The nonlocal phase response $\phi_\mathrm{BG}$ is created by allowing for an angle-dependent refractive index of the media $n(\theta)$, for which we solve to find
\begin{align}
\frac{n(\theta_{\mathrm{SP}})}{n_{\mathrm{BG}}} & =\frac{C\pm\sqrt{C^{2}+(\mathscr{R}^{2}-C^{2})(1+\mathscr{R}^{2}\tan^{2}\theta_{\mathrm{SP}})}}{(1+\mathscr{R}^{2}\tan^{2}\theta_{\mathrm{SP}})\cos\theta_{\mathrm{SP}}},\label{EQ:n_vs_theta}
\end{align} 
where  $\theta_{\mathrm{SP}}$ is the $k$-vector angle  within the spaceplate medium and $C=(\phi_{\mathrm{G}}+2\pi m(\theta_{\mathrm{SP}}))/\phi_\mathrm{BG}(0,d)$ (see \SI{SEC:Nvstheta_derivation} for details). Such a homogeneous non-isotropic plate acts as a spaceplate with compression factor $\mathscr{R}$ for a background medium with refractive index $n_{\mathrm{BG}}$. 

We now discuss the requisite index profile in more detail and identify a physically realizable solution. From here on, we assume $m=0$ for all angles and take the positive root. Since the global phase offset $\phi_{\mathrm{G}}$ is still arbitrary, so is $C$. Thus, $C$ parametrizes an infinite family of solutions, some of which are shown in Fig.~\ref{Fig:schematic}g. One realizable homogeneous solution is with $C=0$  (Fig.~\ref{Fig:schematic}f). This solution is related to but is distinct from a solution from transformation optics that compresses the full electromagnetic field~\cite{Leonhardt2006,Pendry2006,Roberts2009}. Other solutions are presented in \SI{SEC:other_solutions}. Remarkably, we find that the refractive index described by this solution is that of a negative uniaxial birefringent medium ($n_{\mathrm{o}}>n_{\mathrm{e}}$ for ordinary (o) and extraordinary (e) polarizations) with $n_\mathrm{e}=n_\mathrm{BG}$ and its e-axis along $z$ (see \SI{SEC:Nvstheta_derivation} for details). A light field with e-polarization propagating through this medium experiences a compression factor along $z$ of ${\mathscr{R}=(n_{\mathrm{o}}/n_{\mathrm{e}})}$ relative to propagation in isotropic medium $n_\mathrm{BG}$. Theoretically, this negative uniaxial medium acts as a perfect spaceplate for all incident angles.

In order to show conclusively that the spaceplate concept does work in practice and to explore its limitations, we experimentally test this uniaxial spaceplate. (We present tests of a second type of homogeneous spaceplate, a low-index medium, in \SI{SEC:low-index_spaceplate}.) Naturally available uniaxial crystals have $n_{\mathrm{e}}>1$ and so, instead of comparing to propagation in vacuum, we are limited to comparing to a background medium with  $n_\mathrm{e}=n_\mathrm{BG}$, here linseed oil ($n_{\mathrm{BG}}=1.48$). We use a $d=29.84$-mm-long calcite crystal (CaCO$_{3}$) plate with its optic axis oriented perpendicular to its entrance and exit faces. With $n_\mathrm{o} = 1.660$ and $n_\mathrm{e} = 1.486$, the resulting compression factor is a modest $\mathscr{R}=1.12$, far from being of practical use but sufficient for proof-of-principle tests. We propagate a focusing beam through the oil and compare it to the same beam when propagating through the uniaxial spaceplate placed in oil. An ideal spaceplate will shift this focus by $\Delta\equiv d-d_{\mathrm{eff}}=-(\mathscr{R}-1)d$. Looking at Fig.~\ref{Fig:beams_measurement}a, we see that the addition of the spaceplate clearly shifts the focus towards the plate. The measured shift for the e-polarized beam is $\Delta=-3.4$~mm, which agrees well with the predicted shift of $\Delta=-3.5$~mm (see \SI{SEC:polarization_calcite} for details on the o-polarized beam). The spaceplate advances the focus of a beam, just as if it had passed through an additional length of the background medium, thereby showing that the theoretical concept works in practice.

We next experimentally investigate the transverse displacement of a beam incident on a uniaxial spaceplate. This effect is central to the spaceplate's application since it also applies to rays in the standard ray-tracing-based design of lens systems. In order to test whether a uniaxial spaceplate is inherently limited in numerical aperture, we vary the angle of incidence  $\theta$ of the beam with respect to the normal of the calcite interface. For each angle, we record the beam's lateral displacement (indicated by $\Delta x$ in Fig.~\ref{Fig:schematic}a) upon exiting, shown in Fig.~\ref{Fig:beams_measurement}b. The observed displacement $\Delta x$ (red data) is equal to the ideal displacement of a beam travelling through $d_{\mathrm{eff}}$ of the linseed oil at angle $\theta$ (dashed theory curve). Consequently, the uniaxial is found to perfectly reproduce the free-propagation displacement for all measured angles, \emph{i.e.}, up to $\theta=35^{\circ}$, corresponding to $\mathrm{NA}=0.85$ in oil. 

\begin{figure}[!b]
 \centering \includegraphics[width=\textwidth]{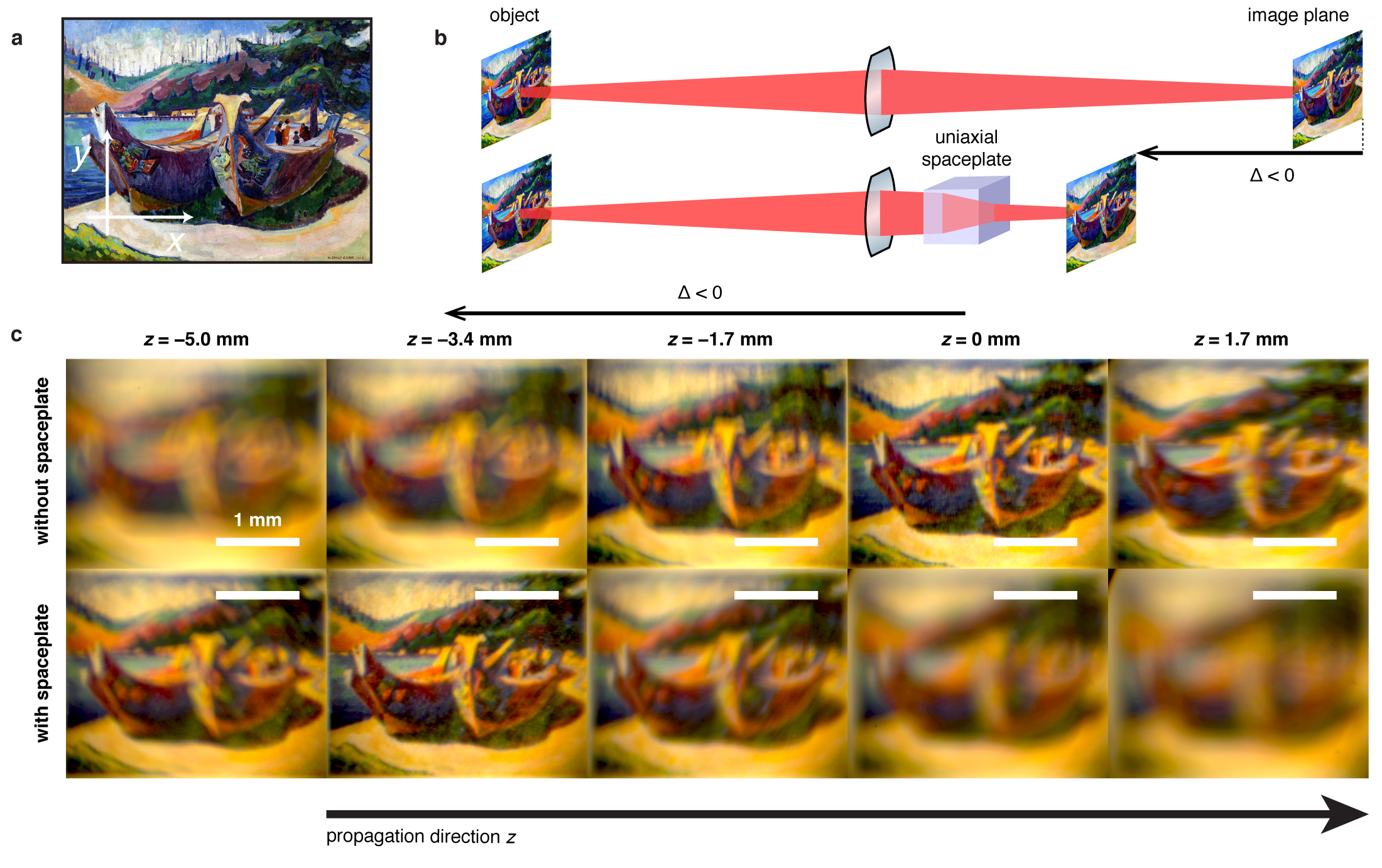}
\caption{\textbf{~|~Advance of a broadband visible image using a spaceplate. a,~}This print of a painting was illuminated with incoherent white light. \textbf{b,~}An image of the print is formed in either a background medium, glycerol, or through the calcite spaceplate in glycerol. \textbf{c,~}Camera images at various distances $z$. The spaceplate advances the focal plane of the image by $\Delta=-3.4$~mm relative to the glycerol alone. The scale bar is the same length in all the images. This result illustrates that the spaceplate does not change the magnification \emph{i.e.,} it does not introduce any focusing power. Note that the yellow tint of the recorded images is due to the illumination.}
\label{Fig:imaging}
\end{figure}

Recently introduced exotic optical material responses, such as negative or near-zero epsilon, have sometimes been associated with material resonances that limit the bandwidth of corresponding devices~\cite{Vulis2018}. To probe whether space compression is an inherently narrowband phenomenon, we test the capability of a spaceplate to reduce the size of a complete full-spectrum visible imaging system. A print of the painting in Fig.~\ref{Fig:imaging}a is illuminated using an incoherent visible white-light source.  A lens system forms an in-focus image of the print at an image plane inside a tank of glycerol placed after the last lens (Fig.~\ref{Fig:imaging}b). While glycerol ($n_{\mathrm{BG}}=1.4743$ ) matches $n_\mathrm{e}$ slightly worse than linseed oil, it has a higher transmittivity across the visible spectrum, which makes it more suitable for full-colour imaging. (See \SI{SEC:Methods}). Figure~\ref{Fig:imaging}c shows images captured by a CCD camera at a series of different positions $z$ along the system axis (see \SI{SEC:visualizations}). At $z=0$, Fig.~\ref{Fig:imaging}c shows that the captured image is in sharp focus, whereas at position $z=-3.4$~mm, the captured image is still out of focus, as it has not propagated far enough to fully form.  We now look at how the spaceplate affects this image formation by placing the calcite crystal into the glycerol before the image plane. The bottom row of Fig.~\ref{Fig:imaging}c shows the images captured at the same $z$ positions as the top row, now with the spaceplate in place. Now, the image comes into focus sooner than with the glycerol alone. Specifically, the captured image is sharp at position $z=-3.4$~mm, whereas in the top row, the image is still forming. Thus, we observe an image advance of $\Delta=-3.4$~mm, in approximate agreement with the theoretical prediction of  $\Delta=-3.5$~mm. The entire color image remains in focus simultaneously, illustrating the broadband operation of the uniaxial spaceplate. Furthermore, the magnification of the image is preserved, as evidenced by comparing the sizes of the images at their respective focal planes. Thus, the lens system has been shortened without changing the field of view, the NA, or the magnification. In contrast, shortening the lens system by reducing the lens focal lengths would change all three of these important imaging system parameters.

\subsection*{Discussion}
We have introduced the concept of the spaceplate and presented two types, which we have simulated and experimentally tested in order to address select potential inherent limitations. The uniaxial spaceplate (and the low-index spaceplate, see \SI{SEC:low-index_spaceplate}) experiments unambiguously demonstrate that a spaceplate is something physically realizable, validating our predictions and Eq.~\ref{EQ:n_vs_theta}. The uniaxial spaceplate experiments, in particular, also show that a spaceplate can be simultaneously broadband in the visible regime, achromatic, have a high NA, as well as high transmission-efficiency, albeit for a small compression factor ($\mathscr{R} = 1.12$). The metamaterial and low-index spaceplate show that a spaceplate may be polarization-independent. Furthermore, the metamaterial spaceplate shows that a spaceplate may have a compression factor that is many times larger than unity ($\mathscr{R} \approx 5$) and not bounded by the ratio of the indices of any of its constituent materials. It remains to be established whether these properties could be combined in a single spaceplate design that has a usefully large compression factor $\mathscr{R}$, similar to what has been accomplished over the last decade in metalenses~\cite{Yu2011,Zheng2015,Khorasaninejad2016,Khorasaninejad2016a,Chen2018,Zhang2018,Shrestha2018}.

In order to gauge the amount of spaceplate improvement that is yet required, we estimate some performance parameters for a few potential applications. We will assume the use of a multilayer spaceplate and, thus, a total thickness $d$ of 100~\micron, the limit of current thin-film coating technology. The first application is inside a modern smartphone camera, which contains typical spacings of $d_\mathrm{eff}=1$ to $4$~mm. A corresponding $\mathscr{R}$ of 10 to 40, only a factor of two to eight more than what we have presented, would compress these distances to $d=100$~\micron, inconsequentially small for a camera. Such a spaceplate would need to operate over the full visible wavelength range, have an NA of 0.2, and be polarization insensitive. All of these properties have been demonstrated in the present work, albeit separately. Moreover, sequentially depositing such a spaceplate and a metalens on top of an image sensor would compress an entire camera into an ultra-thin monolithic form-factor. A second application is to shorten the $d_\mathrm{eff}=2$ to $4$~cm distance between the lens and display in virtual reality (VR) headsets. This distance has been identified by industry as a key obstacle to adoption~\cite{Oculus2018}. An $\mathscr{R}$ of 200 to 400 (for $d=100$~\micron) would effectively eliminate this distance. While the required NA is high (\emph{e.g.,} $>0.6$, but still below that of our uniaxial spaceplate), other requirements are relaxed. In particular, since liquid crystal displays emit only a single polarization and need only emit three narrowband colours (RGB, such as in laser-based displays), the spaceplate need only function correctly for a single polarization and three optical wavelengths. Lastly, in integrated optics, narrowband (\emph{i.e.}, single-wavelength, 1550 nm) single-polarization devices are common~\cite{Halir2016, Cheben2018}. Such a spaceplate, with a modest compression factor of $\mathscr{R}=5$, similar to our multilayer design, would significantly increase the density of devices on a chip. Although improvement in spaceplate performance will be necessary, the requirements of these potential applications suggest that these advances are not so large as to be implausible.

We now consider whether and by what means the compression ratio  $\mathscr{R}$  could be improved. As our metamaterial design is based on optimization methods, it cannot directly inform us on whether or not fundamental physics may impose limits on $\mathscr{R}$. However, causality does not seem likely to constraint $\mathscr{R}$ since the overall time delay for the imaging light to pass through the spaceplate is unconstrained (see \SI{SEC:Causality} for further discussion). Moreover, temporal effects, such as frequency chirp, are free to occur since they are irrelevant to imaging, which is a quasi-static process. Therefore, we do not foresee any fundamental limits to $\mathscr{R}$; however, there may exist fundamental trade-offs between $\mathscr{R}$ and other performance parameters (\emph{e.g.,} operation bandwidth or numerical aperture). Establishing the exact nature and subsequent consequences of these potential trade-offs is a topic for future work.

We propose the following avenues by which the compression factor $\mathscr{R}$ could be increased. First, the feasibility of two of the homogeneous spaceplate solutions amongst the set given by Eq.~\ref{EQ:n_vs_theta} suggests that other solutions in the set may be physically realizable and might bear a higher $\mathscr{R}$ (see \SI{SEC:other_solutions} for a discussion of other solutions). Second, industry has sophisticated design methods that could optimize a multilayer spaceplate made of a stack of thousands of layers of multiple materials, which might yield greatly higher values of $\mathscr{R}$.  A third avenue is engineering an artificial uniaxial medium analogous to the calcite we used. For example, composite materials can be used to create record-breaking anisotropic responses~\cite{Niu2018a}.  Or, as another example, a uniaxial medium can be fabricated by alternating sub-wavelength-thick layers between two materials~\cite{Cai2010,Poddubny2013,Wei2014,Popov2016a}. The latter would create a uniaxial spaceplate with the potential advantages that the birefringence (and, hence, $\mathscr{R}$) can be larger, broadband, and also slowly varied along $z$ to avoid reflection at the interfaces. By following these avenues and others, we expect spaceplates to rapidly improve.

From a broader perspective, the spaceplate further demonstrates the power of nonlocal optical elements that operate directly on the phase of transverse Fourier components of a light field. To the best of our knowledge, this work is the first to design a metamaterial that directly manipulates the phase in $k$-space. Achieving full nonlocal control (\emph{e.g.,} combining the control of transmittance with phase control) would enable all of the benefits of Fourier optics (\emph{e.g.,} spatial filtering) without needing a lens system to access the farfield. In turn, repeatedly iterating between this momentum-dependent Fourier control and position-dependent control has been shown to enable fully arbitrary and lossless spatial-transformations of light fields~\cite{Morizur2010}. Using nonlocal metamaterials and local metasurfaces to respectively accomplish these two controls opens the possibility of complete spatial control of light in a monolithic device.

\section*{Methods}

\subsection*{Homogeneous spaceplates}

\paragraph*{Background medium:} The background medium for measurements in Fig.~\ref{Fig:beams_measurement} is linseed oil (also known as flaxseed oil). This oil (refractive index $n_{\mathrm{BG}}=1.4795$ at an optical wavelength of $\lambda=532$~nm) was chosen to match to $n_\mathrm{e}$ of the uniaxial spaceplate material ($n_\mathrm{e}=1.486$). The colour imaging measurements in Fig.~\ref{Fig:imaging} instead used a background medium of glycerol ($n_{\mathrm{BG}}=1.4743$ at 532~nm). While glycerol matches $n_\mathrm{e}$ slightly worse than linseed oil, it has a higher transmittivity across the visible spectrum, which makes it appropriate for full-colour imaging.

\paragraph*{Uniaxial spaceplate:} We use a 20.04~mm $\times$ 19.98~mm $\times$ 29.84~mm (width $\times$ height $\times$ depth, $\pm 0.06$~mm) right rectangular prism made of calcite that was cut with its extraordinary optical axis along the depth direction. The surfaces perpendicular to this axis are polished and used as the entrance and exit faces. Note that the surface quality is low, which somewhat distorts and scatters the beam in the measurements in Fig.~\ref{Fig:beams_measurement} (see \SI{SEC:physical_spaceplates}). Calcite is negative uniaxial with refractive indices $n_\mathrm{e}=1.486$, $n_\mathrm{o}=1.660$ at a wavelength of $\lambda=532$~nm. For e-polarized light in a background medium with $n_{\mathrm{BG}}=n_\mathrm{e}$, this crystal gives an expected enhancement factor of $\mathscr{R}=n_\mathrm{o}/n_\mathrm{e}=1.117$ and an advance $\Delta=(1-\mathscr{R})d=-3.494$~mm (\emph{i.e.,} a shift towards the crystal). The o-polarized light will experience a medium of isotropic refractive index $n_\mathrm{o}$. Consequently, $R=n_\mathrm{BG}/n_\mathrm{o}=0.895$ and $\Delta=(1-\mathscr{R})d=3.126\thinspace\mathrm{mm}$. 

\paragraph*{Low-index spaceplate:} Our implementation of a low-index spaceplate consists of a glass-faced cylindrical cell containing air (length $d=4.37\pm0.06$~mm and $\mathrm{diameter}=25.82\pm0.06$~mm). The faces are $0.14\pm0.01$~mm thick microscope coverglass pieces (see \SI{SEC:physical_spaceplates}).

\subsection*{Experimental setup}\label{SEC:setup}
The experimental setup is shown in Fig.~\ref{Fig:setupimaging}.

\paragraph*{Light sources:} The measurements in Fig.~\ref{Fig:beams_measurement} used a 4.5~mW diode laser with an optical wavelength of 532~nm. We spatially filter the laser beam with a single-mode fiber. The beam that is then focused to a waist of $\omega_0=9.5\pm0.5$~\micron{} ($e^{-2}$ radius) with a $0.69^{\circ}\pm0.03^{\circ}$ $e^{-2}$ half-angle and a Rayleigh range of $0.79\pm0.08$~mm, all in the background oil medium. The measurements in Fig.~\ref{Fig:imaging} instead used incoherent visible white-light illumination.

\paragraph*{Field relay system:}
Both the beam measurements and imaging measurements use a field relay lens system to relay the full $E_\mathrm{BG}(x,y,z)$ electric field profile to a region outside the tank containing the background medium and spaceplate. The $f_{1}=100$~mm lens after the tank and the $f_{2}=200$~mm lens are separated by a distance $s_{\mathrm{4F2}}=300$~mm, which constitutes a common lens system known as a 4f system. The resulting magnification is $M=f_{2}/f_{1}=2$. The system relays the field outside $E_\mathrm{out}(x,y,z)$ such that $E_\mathrm{out}(Mx,My,M^2z/n_\mathrm{BG})\propto E_\mathrm{BG}(x,y,z)$.  Outside the tank, we use an image sensor (CCD) to record the intensity spatial-distribution in the $x,y$ plane. We then scan the CCD along $z$. Five images are taken at each step and averaged to reduce camera noise. A shutter is closed in order to acquire background images, which are subtracted from the raw images to compensate for stray light and camera noise. For the measurements in Fig.~\ref{Fig:beams_measurement}, we use a monochromatic camera ($3088\times2076$~pixels, 2.4~\micron{} $\times$ 2.4~\micron{} each, 12 bit). For the measurements in Fig.~\ref{Fig:imaging}, we use a colour camera ($1936\times1216$~pixels, 5.86~\micron{} $\times$ 5.86~\micron{} each, 12 bit). In the figures, we report the dimensions of the field inside the oil.

\paragraph*{Beam measurements:}
The beam measurement setup is shown in Fig.~\ref{Fig:setupimaging}a. A diode laser produces a beam of wavelength 532~nm with a power of 4.5~mW. This beam is attenuated using a filter and then has its spatial mode filtered by a single-mode fiber. The beam exiting the fiber is collimated. A half-waveplate ($\lambda/2$) and polarizing beamsplitter (PBS) are used to vary the beam intensity and polarize the beam. The beam's polarization is subsequently controlled by a zero-order half-waveplate ($\lambda/2$) and quarter-waveplate ($\lambda/4$). The lens before the tank, $f_{1}=100$~mm, is used to focus the beam through the spaceplate. The spaceplate's entrance surface is located 80~mm from this $f_1$ lens. The tank contains linseed oil as a background medium. We then use the field relay system to image the transmitted beam.
For the measurements in Fig.~\ref{Fig:beams_measurement}a, we move the camera along $z$, recording an image at steps of 0.02~in (0.508~mm) over a range of 60~mm. For the measurements in Fig.~\ref{Fig:beams_measurement}b, the camera $z$-position is set so that the camera images the beam focus. In order to measure the lateral beam displacement $\Delta x$, the spaceplate is then tilted by an angle $\theta$ about $y$ in steps of $0.25^{\circ}$ over a range of $40^{\circ}$ and $43.5^{\circ}$ for the calcite and air plates, respectively (note that these ranges are the maximum allowed by the clear aperture of the respective spaceplate). For each camera position $z$ or crystal angle $\theta$, the recorded image is summed along the $y$-direction to arrive at an intensity distribution along $x$. These $x$ intensity distributions are presented along the vertical direction of the plots in Fig.~\ref{Fig:beams_measurement}. 

\paragraph*{Imaging measurements:}
The imaging measurement setup is shown in Fig.~\ref{Fig:setupimaging}b. With visible white light, we illuminate a 15~mm $\times$ 12~mm print of a painting (\textit{First Nations War Canoes in Alert Bay} by Emily Carr, 1912) printed on ordinary white paper. At a distance $s_{oi}=475$~mm from the print is a lens of focal length $f_{i}=500$~mm. A further $s_{i1}=355$~mm from the $f_i$  lens is the first $f_{1}=100$~mm lens. Together this lens pair (NA$=0.025$) collects the the light reflected from the print, transmits the light through the spaceplate in the tank, which then forms in the background medium an image of the print with magnification $0.209\pm0.001$. Between this lens pair is a linear film polarizer (visible broadband, $400-700$~nm), which we rotate to set the polarization of the light. The spaceplate is placed $s_{\mathrm{SP}} \approx 80$~mm after this $f_1$ lens. The background medium in the tank is now glycerol rather than linseed oil since the former has a high transmission across the visible spectrum, which is ideal for full-colour imaging. We then use the field relay system to image the field after the spaceplate at various propagation distances $z$. For the measurements in Fig.~\ref{Fig:imaging}, we move the camera along $z$, recording an image at steps of 0.02~in (0.508~mm) over a range of 100~mm. In the focused image plane, the image has $x\times y$ dimensions of 6.25~mm $\times$ 5.03~mm on the camera sensor.

\paragraph*{Polarization control:} A uniaxial spaceplate acts to replace space for e-polarized light. However, the e-polarization direction varies depending on the angle of the incident wavevector relative to the crystal's extraordinary optic axis. In order for the incident light field to be simultaneously e-polarized and approximately uniformly polarized along one direction, the crystal is tilted slightly about $y$ by an angle $\alpha$ relative to the incident beam (and system axis). The tilt is $\alpha=4.5^{\circ}$ for Fig.~\ref{Fig:beams_measurement}a and $\alpha=8^{\circ}$ for Fig.~\ref{Fig:imaging}. An $x$-polarized light field will then be e-polarized with respect to the crystal; a $y$-polarized field will be o-polarized. For the laser, the incident polarization is set by a polarizing beamsplitter followed by waveplates. The polarization of the white light is set by a film polarizer designed for broadband visible light. More generally, the uniaxial spaceplate works for a lightfield with an angularly non-uniform polarization that is extraordinary everywhere (\emph{e.g.,} a radially polarized field).

\paragraph*{Coordinate system:} We use $x\times y\times z$ as a coordinate system for the experiment, where $x$ and $y$ are the transverse directions and $z$ is the optical system axis (\emph{i.e.,} the beam axis). The crystal's height dimension is along $y$. The extraordinary optical axis of the tilted crystal defines $z'$ of a second coordinate system, $x'\times y\times z'$. Thus, the second coordinate system is related to the first by a rotation about $y$ by angle $\alpha$. The uniaxial spaceplate always acts as $d_{\mathrm{eff}}=\mathscr{R}d$ distance along $z'$ with $\mathscr{R}=n_\mathrm{o}/n_\mathrm{e}$. This tilt reduces the effective distance along $z$ by a factor $\cos\alpha$, which for small $\alpha$ is approximately unity.

\subsection*{Multilayer metamaterial spaceplate}

\paragraph*{Metamaterial structure:} We consider structures made up of planar layers alternating between two materials, silicon and silica (\emph{i.e.,} a `multilayer stack'). Each layer can have an arbitrary thickness larger than 10~nm, set by feasible fabrication capabilities. The combined thickness of the entire stack is designed to be approximately 10~\micron.

\paragraph*{Genetic algorithm:} Our aim is to design a multilayer stack to replace a background medium of vacuum. To do so, we search for a structure that gives a phase profile $\phi_{\mathrm{SP}}$ that matches the phase profile  $\phi_\mathrm{BG}(d_{\mathrm{eff}})$ resulting from propagation through a slab of vacuum of length $d_{\mathrm{eff}}$. We restrict this aim to a range of incident angle from zero to $\theta_{\mathrm{max}}$ (\emph{i.e.,} the NA of the spaceplate). The search is conducted with a genetic algorithm whose goals are to maximize $d_\mathrm{eff}$ while minimizing any optical aberration resulting from a non-ideal phase profile. To quantify the latter goal, we first calculate the difference of the slope from that of the ideal profile, $\Delta\phi'=\phi'_\mathrm{SP}-\phi'_\mathrm{BG}$, where $\phi'=\partial\phi/\partial \theta$. This angular slope is the relevant quantity to consider since any global phase $\phi_{G}$ and phase wraps $2\pi m$ will be eliminated by the derivative. We then find the root-mean-square (RMS) of this difference, $\Delta\phi'_{\mathrm{RMS}}$. The RMS deviation $\Delta\phi'_{\mathrm{RMS}}$  is an optical aberration that results in an increased beam waist $\omega_{\mathrm{SP}}=\omega_0(1+\theta_{\mathrm{max}}\Delta\phi'_{\mathrm{RMS}})$ relative to the waist $\omega_0=\lambda/(\pi\theta_\mathrm{max})$ in the absence of the multilayer stack. As a worst case scenario, this larger waist will increase the Rayleigh range to $z_{SP}=\pi\omega_{\mathrm{SP}}^2/\lambda$.  The parameter $z_\mathrm{SP}$ increases with aberration and the inverse of the usable angle $\theta_{\mathrm{max}}^{-1}$ . The two goals of the algorithm can be combined in a single fitness function, $F=d_{\mathrm{eff}}/z_{SP}=\pi d_{\mathrm{eff}}\theta_{\mathrm{max}}^2/(\lambda(1+\theta_{\mathrm{max}}\Delta\phi'_{\mathrm{RMS}})^2)$, where we have used the small-angle approximation repeatedly. The larger the value of $F$ is, the better the performance of the multilayer spaceplate will be.

We now outline the functioning of the genetic algorithm. Each generation in the genetic algorithm had a population size of 500. The DNA of each population member was the material and the thickness of each layer in the stack. We used two materials, silica and silicon. The maximum number of layers was set to 40 and each layer was constrained to have a thickness greater than 10~nm. For each member, we use the standard transfer matrix formalism to calculate the complex transmission amplitude $H=|H|\exp(i\phi_\mathrm{SP})$ of the multilayer stack for a set of incident angles $\theta$.  We use nonlinear regression to fit $\phi_\mathrm{SP}$ with an ideal phase profile $\phi_\mathrm{BG}(d_{\mathrm{eff}})$, giving $d_\mathrm{eff}$ and, with this fit, we numerically calculate $\phi'_{\mathrm{RMS}}$. Both the fit and calculation are conducted over a range of input angles from zero to $\theta_{\mathrm{max}}=15^\circ$. With these performance parameters we find the fitness $F$ of each population member. The device thickness of the first generation is constrained to 10~\micron, but this parameter is not constrained for later generations. The algorithm was carried out until there was a convergence in the fitness of the ``best'' member of each generation. For the structure reported here, this took 4000 generations.

\paragraph*{Full-wave simulations:} The simulation in Fig.~\ref{Fig:metamaterial} was performed using a commercial 2D finite-difference time-domain solver. The boundary conditions are perfectly-matched layers. Exact details about the geometry of the structure and material parameters can be found in \SI{SEC:metamaterial}.

\paragraph{Acknowledgements} The authors acknowledge support from the Transformative Quantum Technologies program of the Canada First Research Excellence Fund, the Canada Research Chairs Program, and the Natural Sciences and Engineering Research Council of Canada. AA was supported by Mitacs Globalink. OR acknowledges the support of the Banting Postdoctoral Fellowship of NSERC. The authors thank Eric Mazur for suggesting the use of glycerol. While our paper was in review, another paper was posted to the arXiv preprint server which theoretically demonstrates a spaceplate based on a 2D photonic crystal~\cite{Guo2020}.

\paragraph{Author Contributions} OR, JL, and RB conceived the basic idea for this work. JL did the theory work. OR, KB, AA, JL, and LG designed the experiment. MD and OR carried out the measurements. OR analysed the experimental results. OR, AA, and JL did the multilayer design. OR performed the full-wave simulations. JL and RB supervised the research and the development of the manuscript. OR wrote the first draft of the manuscript, and all authors subsequently took part in the revision process and approved the final copy of the manuscript.

\clearpage

\onecolumn
\renewcommand\thepage{S\arabic{page}} 
\setcounter{page}{1}
\renewcommand\thesection{S\arabic{section}} 
\setcounter{section}{0}
\renewcommand\thefigure{S\arabic{figure}}   
\setcounter{figure}{0}  
\renewcommand\theequation{S\arabic{equation}} 
\setcounter{equation}{0}

\noindent{\Huge Supplementary Information} \vspace{1em}

\noindent Below is the Supplementary Information for \emph{An optic to replace space and its application towards ultra-thin imaging systems} by Orad Reshef, Michael P. DelMastro, Katherine K. M. Bearne, Ali H. Alhulaymi, Lambert Giner, Robert W. Boyd, and Jeff S. Lundeen. In Sec.~\ref{SEC:Methods} we have a figure depicting the experimental setup. In Sec.~\ref{SEC:Telephoto}, we compare the operation of a spaceplate to that of a telephoto lens. Section~\ref{SEC:metamaterial} summarizes details about the nonlocal metamaterial spaceplate structure. In Sec.~\ref{SEC:Nvstheta_derivation}, we derive the refractive index for an anisotropic homogeneous spaceplate medium. We then step through the different types of solutions that are yielded by this derivation and their properties. In Sec.~\ref{SEC:low-index_spaceplate} we present the low-index spaceplate, including experimental validation. In Sec.~\ref{SEC:polarization_calcite}, we describe ordinarily and extraordinarily polarized beam measurements conducted with the uniaxial spaceplate, including additional beam-focus measurements as proof of the two-dimensional action of the spaceplate. Section~\ref{SEC:visualizations} contains the captions for the movies. In Sec.~\ref{SEC:Causality}, we have included a discussion on the implications of causality on the inherent limitations of a spaceplate. In Sec.~\ref{SEC:physical_spaceplates}, we describe the fabricated spaceplates we use in our measurements. Finally, in Sec.~\ref{SEC:Shift_model}, we derive the lateral shift $\Delta x$. 

\section{Experimental Setup}\label{SEC:Methods}
\begin{figure}[H]
    \centering
    \subfloat[]{{ \includegraphics[width=0.45\linewidth]{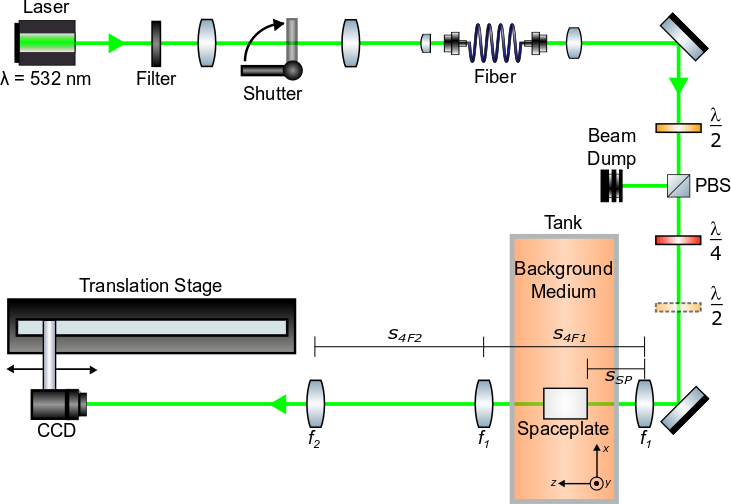} }}
    \qquad
    \subfloat[]{{ \includegraphics[width=0.45\linewidth]{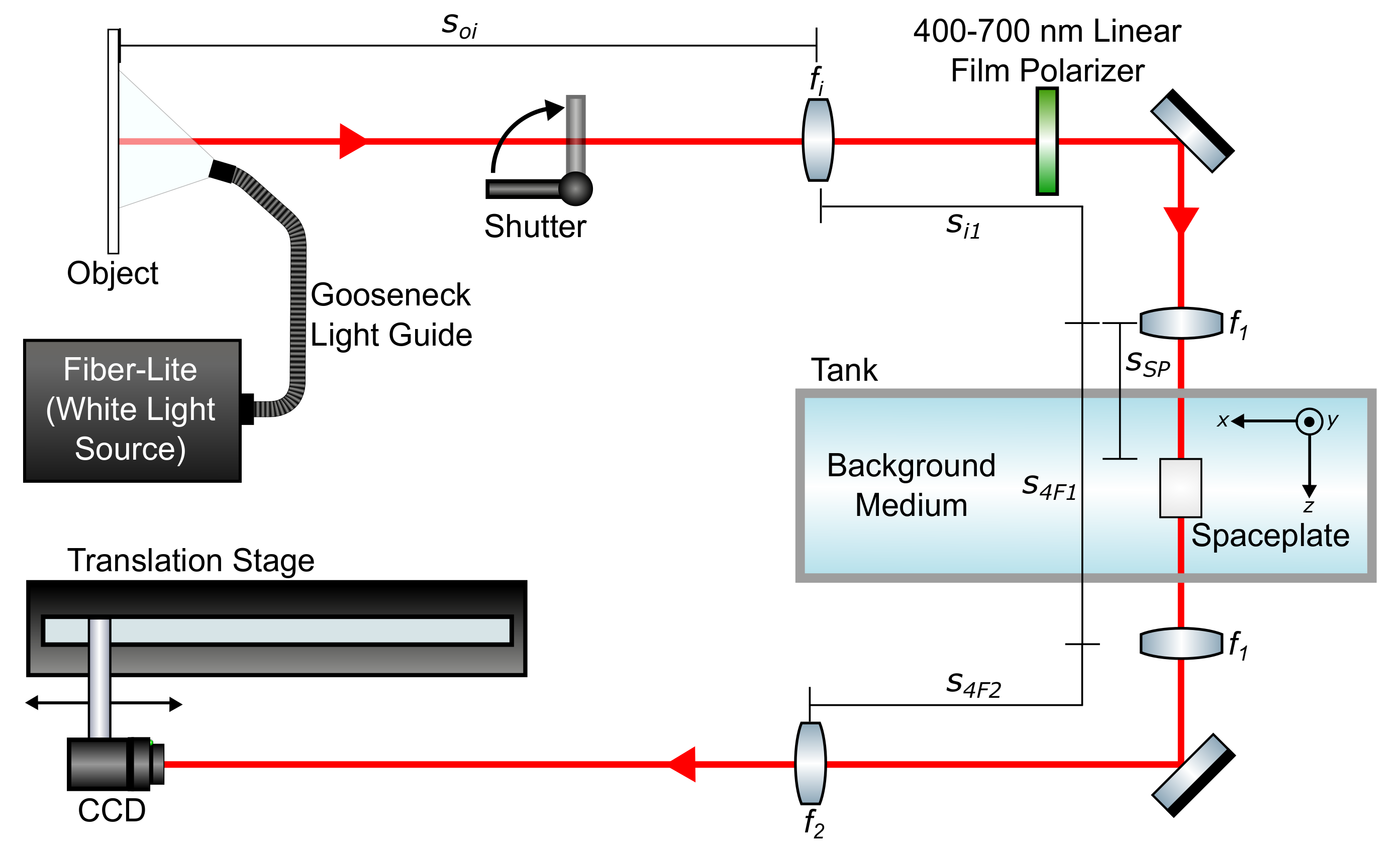} }}
    \caption{\textbf{~|~Experimental setup to measure the action of a spaceplate.} See the text for details. \textbf{a,~}We focus a beam through a spaceplate and measure its shift in $z$ and $x$. \textbf{b,~}We perform full colour imaging through a spaceplate.}
    \label{Fig:setupimaging}
\end{figure}

Depicted in Fig.~\ref{Fig:setupimaging} are the setups for the beam measurements (Fig.~\ref{Fig:setupimaging}a) and the imaging measurements (Fig.~\ref{Fig:setupimaging}b). Details can be found in the Methods section.

\section{Comparison to a telephoto lens}\label{SEC:Telephoto}
\begin{figure}[H]
    \centering
 \includegraphics[width=80mm]{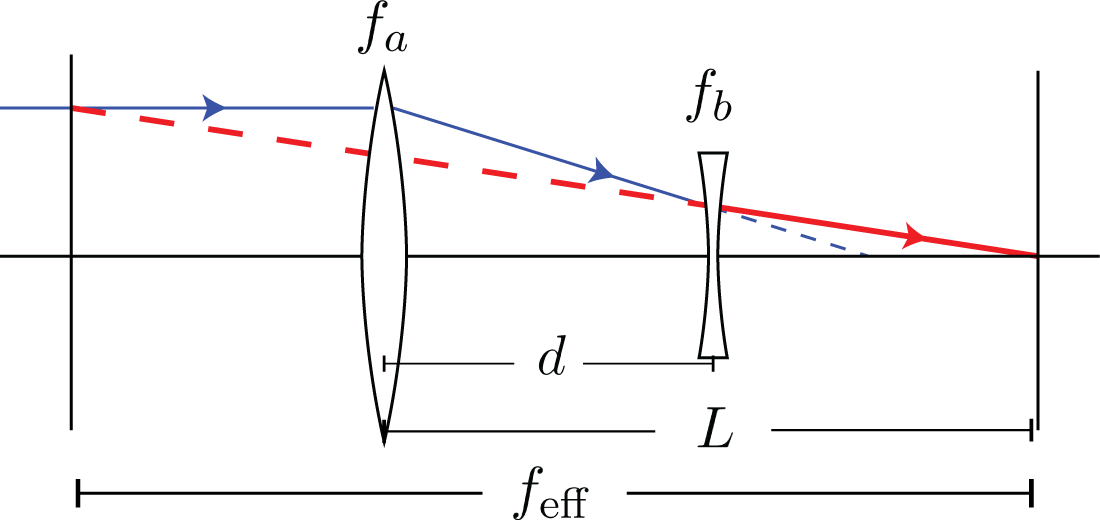}
    \caption{\textbf{~|~A telephoto lens.} }
    \label{Fig:telephoto}
\end{figure}

We compare the properties of a spaceplate with another familiar optical element that is used to reduce the size of an imaging systems--- a telephoto lens. A traditional telephoto lens comprises two components: a converging (positive) lens and a diverging (negative) lens, separated by a distance $d$ (Fig.~\ref{Fig:telephoto}). This combined system decouples the effective focal length from the working distance or the back focal length of the lens. There are established methods for engineering a lens system to exhibit a given effective focal length, with the ratio between the entire track length $L$ and the effective focal length $f_\mathrm{eff}$ defined as the telephoto ratio $k=L/f_\mathrm{eff}$. It is in principle possible to obtain any telephoto ratio if lenses can be aberration-free, flat and thin, and have any focal length and diameter. However, practical considerations (such as the minimum focal length and alignment between lenses) limit the practical values of $k\sim 0.8$~\cite{Kingslake2010}. 

A second point to consider is that the propagation $d$ between the two lenses is critical to the operation of a telephoto; therefore, even for a specific telephoto lens with a vanishing working distance, this required propagation length imposes a minimum size on the system. The spaceplate concept provides an avenue with which one may completely eliminate this space. In fact, a spaceplate could be included within a telephoto lens to reduce its size, for example.

Finally, we note that though a system comprising both a lens and a spaceplate performs a function similar to a telephoto lens, the spaceplate on its own performs a unique function. In particular, as opposed to a telephoto lens, which needs to be designed with \emph{a priori} knowledge of all of the focal lengths, a spaceplate could be added after the fact. It could also be used on its own in other applications that do not fit a telephoto lens system to reduce arbitrary propagation lengths, as it has no lens power and therefore adds no undesired magnification to the system. 

\section{Spaceplate metamaterial} \label{SEC:metamaterial}

\begin{figure}[H]
    \centering
 \includegraphics[width=89mm]{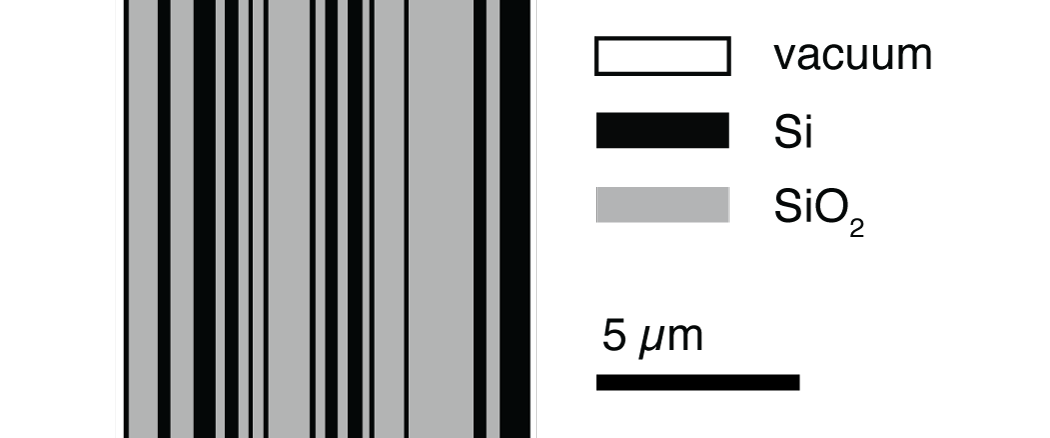}
    \caption{\textbf{~|~Schematic of the nonlocal metamaterial spaceplate.} The spaceplate consists of a multilayer stack formed of two materials, silicon (Si) and silicon dioxide (SiO$_2$). The schematic is to scale for the given scalebar.}
    \label{Fig:metamaterialstructure}
\end{figure}

The metamaterial is designed for operation at an optical wavelength of $\lambda=1550$~nm. At this wavelength, the complex refractive index of silicon is $n_\mathrm{Si}=3.48985+0.00982674i$, and the refractive index of silica glass is $n_\mathrm{SiO_2}=1.45611$. The device has 25 layers and a total thickness of 10.1752~\micron. Table~\ref{TAB:layers} lists the individual layer thicknesses; Figure~\ref{Fig:metamaterialstructure} depicts the cross-section of the metamaterial to scale.

\begin{figure}[H]
    \centering
 \includegraphics[width=70mm]{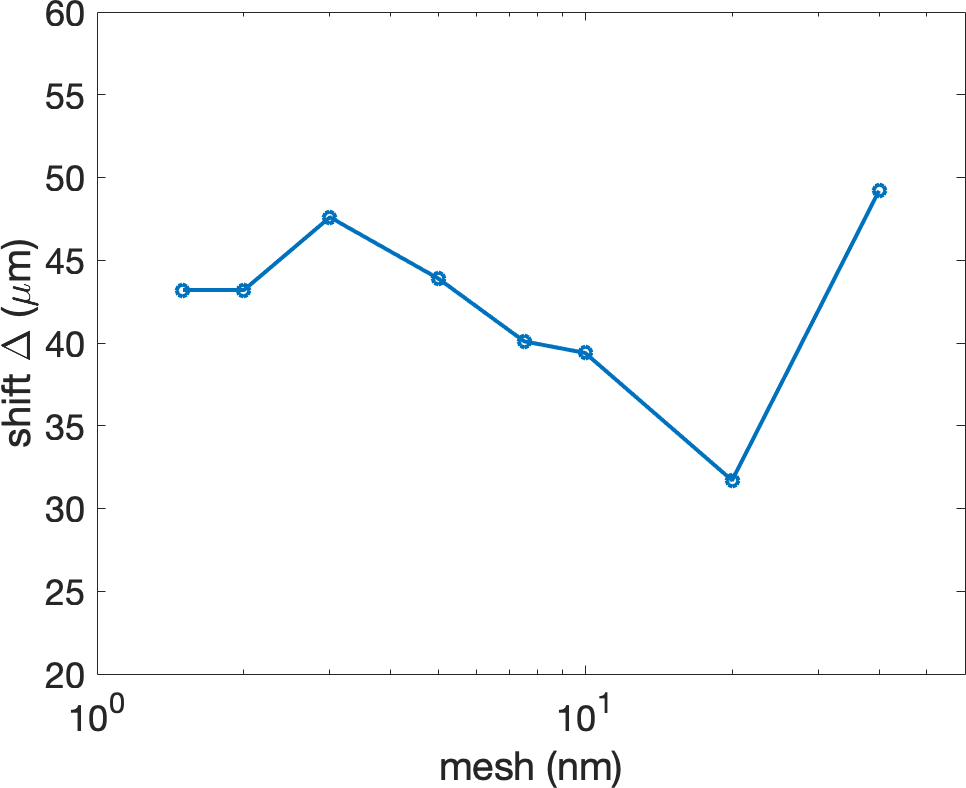}
    \caption{\textbf{~|~Mesh convergence test for the finite-difference time-domain simulation.} The shift of the focus $\Delta$ relative to vacuum is simulated as a function of the smallest mesh size along the direction of propagation within the metamaterial. Here, the smallest mesh used was 2~nm. Critically, all these simulations show a sizeable negative shift and, thus, exhibit the spaceplate effect.}
    \label{Fig:metamaterial_convergence}
\end{figure}

The values in the Table~\ref{TAB:layers} are listed up to 3 decimal places, as was produced by the genetic algorithm.  As can be inferred by the convergence test shown in Fig.~\ref{Fig:metamaterial_convergence}, the device is robust to fabrication imperfections, and so this level of precision is not necessary to generate a metamaterial spaceplate that demonstrates a focus advance.

\begin{table}[H]
\scriptsize
\begin{center}
\begin{tabular}[t]{|c|c|c|}
\hline
Layer & Material & Thickness (nm) \\ \hline
1 & Si & 133.86 \\ \hline
2 & SiO$_2$ & 722.324 \\ \hline
3  & Si &  319.406 \\ \hline
4 & SiO$_2$ & 573.293 \\ \hline
5 & Si & 551.083 \\ \hline
6 & SiO$_2$ & 232.074 \\ \hline
7 & Si & 340.955 \\ \hline
8 & SiO$_2$ & 254.686 \\ \hline
9 & Si & 105.252 \\ \hline
\end{tabular}
\quad
\begin{tabular}[t]{|c|c|c|}
\hline
Layer & Material & Thickness (nm)\\ \hline
10 & SiO$_2$ & 265.555 \\ \hline
11  & Si & 124.592 \\ \hline
12 & SiO$_2$ & 1032.82 \\ \hline
13  & Si & 145.7 \\ \hline
14 & SiO$_2$ & 239.521 \\ \hline
15 & Si & 313.439 \\ \hline
16 & SiO$_2$ & 252.054 \\ \hline
17 & Si & 371.439 \\ \hline
\end{tabular}
\quad
\begin{tabular}[t]{|c|c|c|}
\hline
Layer & Material & Thickness (nm)\\ \hline
18 & SiO$_2$ & 168.605 \\ \hline
19 & Si & 125.935 \\ \hline
20 & SiO$_2$ & 747.517 \\ \hline
21  & Si & 105.681 \\ \hline
22 & SiO$_2$ & 1629.52 \\ \hline
23  & Si & 318.601 \\ \hline
24 & SiO$_2$ & 334.673 \\ \hline
25 & Si & 766.62 \\ \hline
\end{tabular}
\end{center}
\caption{~|~Materials and thicknesses of individual layers in the metamaterial spaceplate shown in Fig.~\ref{Fig:metamaterialstructure}.}\label{TAB:layers}
\end{table}

\begin{figure}[H]
    \centering
 \includegraphics[width=160mm]{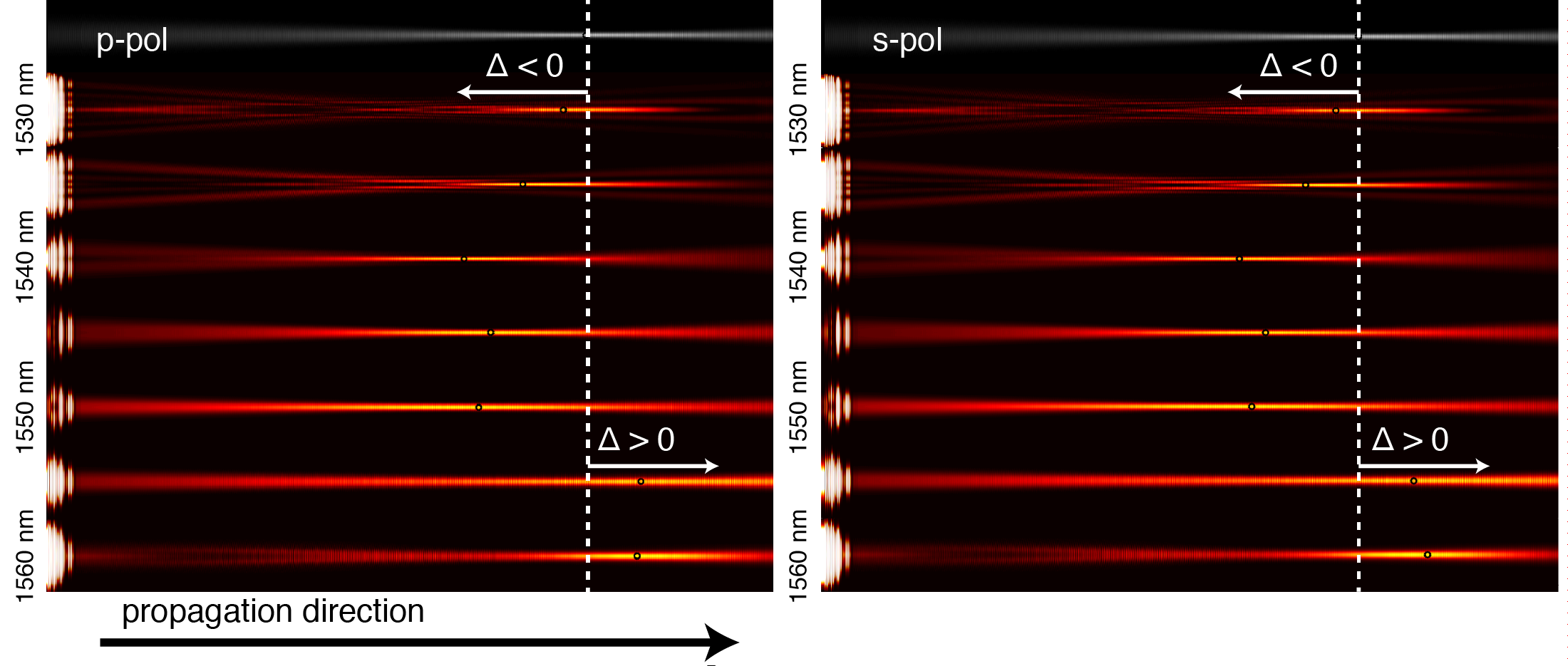}
 
 \includegraphics[width=70mm]{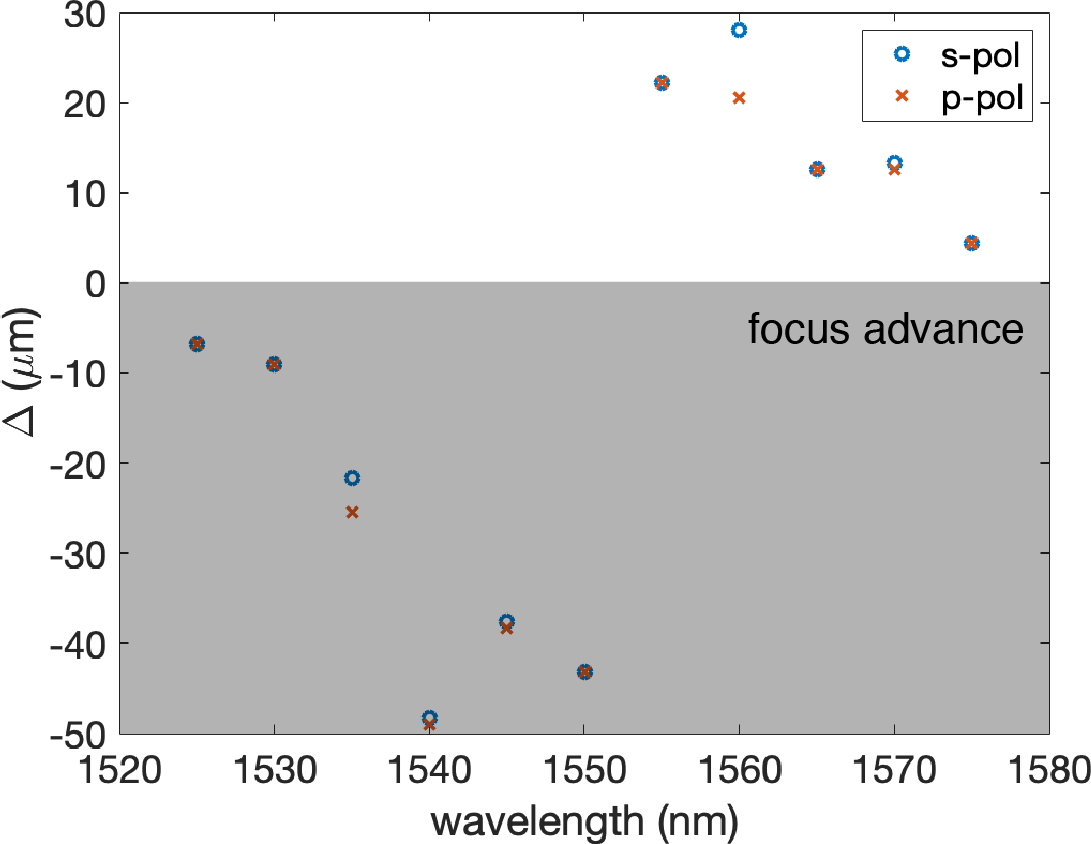}
    \caption{\textbf{~|~Broadband operation of the metamaterial spaceplate. Top,} full-wave simulations of the magnitude of the electric field $|E|^2$ of a focusing Gaussian beam propagating in vacuum (grey, top row) and after propagating through the metamaterial for a range of operating wavelengths (red, all rows but the top row). The device generates a focus advance ($\Delta<0$) for both polarizations and for  wavelengths ranging from 1530~--~1560~nm. \textbf{Bottom,} the total focus advance $\Delta$ as a function of wavelength for this device. }
    \label{Fig:metamaterial_broadbandsim}
\end{figure}

Figure~\ref{Fig:metamaterial_broadbandsim} demonstrates the performance of the metamaterial spaceplate for both p-polarized and s-polarized focusing Gaussian beams for operating wavelengths ranging from 1530~nm to 1560~nm. 

\section{Homogeneous spaceplate solutions}

\label{SEC:Nvstheta_derivation}

\subsection{General solution for an angle-dependent refractive index}

The goal of a spaceplate is to transform an incoming light-field in an
identical manner (for the purposes of imaging) to propagation through a slab of an isotropic homogeneous medium.
To do so, the spaceplate must multiply the complex amplitude of each
plane-wave in an incoming field by the function, $\exp(i\phi_{\mathrm{BG}})$.
In particular, a spaceplate of thickness $d$ must impart a phase $\phi_{\mathrm{SP}}$
that is equal to the phase $\phi_{\mathrm{BG}}$ due to the propagation
through distance $d_{\mathrm{eff}}$ of the medium. Both $d$ and
$d_{\mathrm{eff}}$ are along what we set to be the $z$-axis. 

In more detail, in a medium of index $n_{\mathrm{BG}},$ a plane-wave
has phase difference, $\phi_{\mathrm{BG}}=\mathbf{k}^{(\mathrm{BG})}\cdot\mathbf{r},$
between two positions separated by vector $\mathbf{r}=(x,y,z=d_{\mathrm{eff}})$,
where $\mathbf{k}^{(\mathrm{BG})}=(k_{x}^{\mathrm{(BG)}}\negthickspace,\,k_{y}^{\mathrm{(BG)}}\negthickspace,\,k_{z}^{\mathrm{(BG)}})$
is the wavevector. Here, $|\mathbf{k}^{(\mathrm{BG})}|=n_{\mathrm{BG}}k_{0}\equiv k_{\mathrm{BG}}$
is the wavenumber in the medium and $k_{0}=2\pi/\lambda$ is the vacuum
wavenumber. Inside the spaceplate, the wavevector is $\mathbf{k}^{(\mathrm{SP})}=(k_{x}^{\mathrm{(SP)}}\negthickspace,\,k_{y}^{\mathrm{(SP)}}\negthickspace,\,k_{z}^{\mathrm{(SP)}})$.
If we consider spaceplates made up of plates or layers whose interfaces
are $x,y$ planes, then $k_{x}^{\mathrm{(SP)}}=k_{x}^{\mathrm{(BG)}}$
and $k_{y}^{\mathrm{(SP)}}=k_{y}^{\mathrm{(BG)}}$ due to momentum
conservation. Consequently, such a spaceplate will automatically reproduce
the contribution to the phase due to displacement in $\mathbf{r}$ by
$x$ and $y$, \emph{i.e.,} $\phi_{\mathrm{SP}}=k_{x}^{\mathrm{(SP)}}x+k_{y}^{\mathrm{(SP)}}y+k_{z}^{\mathrm{(SP)}}d=k_{x}^{\mathrm{(BG)}}x+k_{y}^{\mathrm{(BG)}}y+k_{z}^{\mathrm{(SP)}}d$.
The remaining component to be reproduced is the phase due to the displacement along $z$ in the medium,  $\phi_{\mathrm{BG}}=k_{z}^{\mathrm{(BG)}}d_{\mathrm{eff}}$.
In a homogeneous spaceplate this phase is $\phi_{\mathrm{SP}}=k_{z}^{\mathrm{(SP)}}d$.
With this definition, the goal reduces to making the Fourier transfer
function for propagation, $H=\exp(i\phi)$, equal for the spaceplate
and the medium up to a global offset phase $\phi_{\mathrm{G}}$: 
\begin{align}
e^{i\phi_{\mathrm{SP}}} & =e^{i\phi_{\mathrm{BG}}+i\phi_{\mathrm{G}}}\nonumber \\
\implies2\pi m & =\phi_{\mathrm{SP}}-\phi_{\mathrm{BG}}-\phi_{\mathrm{G}}\label{EQ:phaseCondition}\\
2\pi m & =k_{z}^{\mathrm{(SP)}}d-k_{z}^{\mathrm{(BG)}}d_{\mathrm{eff}}-\phi_{\mathrm{G}},
\end{align}
where $m$ is an integer. 

If the plane-wave is traveling in the medium at an angle $\theta$
to the $z$-axis, we can express the $z$-component of the wavevector
as
\begin{align}
k_{z}^{\mathrm{(BG)}} & =k_{\mathrm{BG}}\cos{\theta}.\label{EQ:}
\end{align}
In order to match this $\cos{\theta}$ variation, the spaceplate must
produce an angle-dependent phase. To produce this dependence, one
possible scenario is a spaceplate made of a non-isotropic material.
That is, we consider a spaceplate material with a refractive index
$n(\theta_{\mathrm{SP}})$ that varies with the wavevector angle inside
the spaceplate $\theta_{\mathrm{SP}}$, such that the wavenumber would
be $k_{\mathrm{SP}}=n(\theta_{\mathrm{SP}})k_{0}$. In this case, the
$z$-component of the wavevector is
\begin{equation}
k_{z}^{\mathrm{(SP)}}=n(\theta_{\mathrm{SP}})k_{0}\cos{\theta_{\mathrm{SP}}}=\tilde{n}k_{\mathrm{BG}}\cos{\theta_{\mathrm{SP}}},\label{EQ:SP_wavevector}
\end{equation}
where we defined the refractive index ratio, $\tilde{n}=\tilde{n}(\theta_{\mathrm{SP}})\equiv n(\theta_{\mathrm{SP}})/n_\mathrm{BG}$.
Since transverse momentum is conserved throughout, we define it as
a single parameter $k_{\perp}\equiv\sqrt{(k_{x}^{\mathrm{(SP)}})^{2}+(k_{y}^{\mathrm{(SP)}})^{2}}=\sqrt{(k_{x}^{\mathrm{(BG)}})^{2}+(k_{y}^{\mathrm{(BG)}})^{2}}$.
In particular, by the Pythagorean theorem, $k_{\perp}^{2}=k_{\mathrm{SP}}^{2}-\left(k_{z}^{\mathrm{(SP)}}\right)^{2}$and, also,
\begin{align}
k_{z}^{\mathrm{(BG)}} & =\sqrt{k_{\mathrm{BG}}^{2}-k_{\perp}^{2}}\nonumber \\
 & =\sqrt{k_{\mathrm{BG}}^{2}-\left(k_{\mathrm{SP}}^{2}-\left(k_{z}^{\mathrm{(SP)}}\right)^{2}\right)}\nonumber \\
 & =\sqrt{k_{\mathrm{BG}}^{2}-\tilde{n}^{2}k_{\mathrm{BG}}^{2}\left(1-\cos^{2}{\theta_{\mathrm{SP}}}\right)}\nonumber \\
 & =k_{\mathrm{BG}}\sqrt{1-\tilde{n}^{2}\sin^{2}\theta_{\mathrm{SP}}}.\label{EQ:BG_wavevector}
\end{align}
The goal of these manipulations was to ensure that the only angle that appears in the expressions for both wavevectors is $\theta_{\mathrm{SP}}$. Note that these relations inherently account for refraction at the interfaces.  

We insert the two wavevector equations, Eq.~\eqref{EQ:SP_wavevector}
and Eq.~\eqref{EQ:BG_wavevector}, into the phase condition in Eq.~\eqref{EQ:phaseCondition} to obtain,
\begin{align}
2\pi m=k_{\mathrm{BG}}\,d\,\tilde{n}\,\cos{\theta_{\mathrm{SP}}}-k_{\mathrm{BG}}\,d_{\mathrm{eff}}\,\sqrt{1-\tilde{n}^{2}\sin^{2}\theta_{\mathrm{SP}}}-\phi_{\mathrm{G}}.
\end{align}
We can rearrange this equation to isolate the phase offsets in a single
parameter, 
\begin{align}
C\equiv\left(m+\frac{\phi_{\mathrm{G}}}{2\pi}\right)
\frac{\lambda}{n_{\mathrm{BG}}d}=\tilde{n}\;\cos\theta_{\mathrm{SP}}-\frac{d_{\mathrm{eff}}}{d}\sqrt{1-\tilde{n}^{2}\sin^{2}\theta_{\mathrm{SP}}}.
\end{align}
We then solve for $\tilde{n}$ and recall the definition of $\mathscr{R=}d_{\mathrm{eff}}/d$
to yield Eq.~\eqref{EQ:n_vs_theta} in the main text, the general equation that describes the index of refraction
for a homogeneous spaceplate, 
\begin{align}
\tilde{n}(\theta_{\mathrm{SP}})=\frac{n(\theta_{\mathrm{SP}})}{n_{\mathrm{BG}}} & =\frac{C\pm\sqrt{C^{2}+(\mathscr{R}^{2}-C^{2})(1+\mathscr{R}^{2}\tan^{2}\theta_{\mathrm{SP}})}}{(1+\mathscr{R}^{2}\tan^{2}\theta_{\mathrm{SP}})\cos\theta_{\mathrm{SP}}}.\label{EQ:SI_n_theta}
\end{align}
The physical meaning of $C$ can be understood by re-expressing it
as
\begin{align}
C\left(\theta_{\mathrm{SP}}\right)=\frac{2\pi m\left(\theta_{\mathrm{SP}}\right)+\phi_{\mathrm{G}}}{k_{\mathrm{BG}}d},
\end{align}
where we have made the potential dependence of the integer $m$ on
angle explicit. $C$ is the ratio of the total phase offset to the
phase $\phi_{\mathrm{BG}}$ accumulated by a wave at angle $\theta_{\mathrm{SP}}=\theta=0$,
traveling distance $d$.
\begin{figure*}[htb]
 \centering \includegraphics[width=\textwidth]{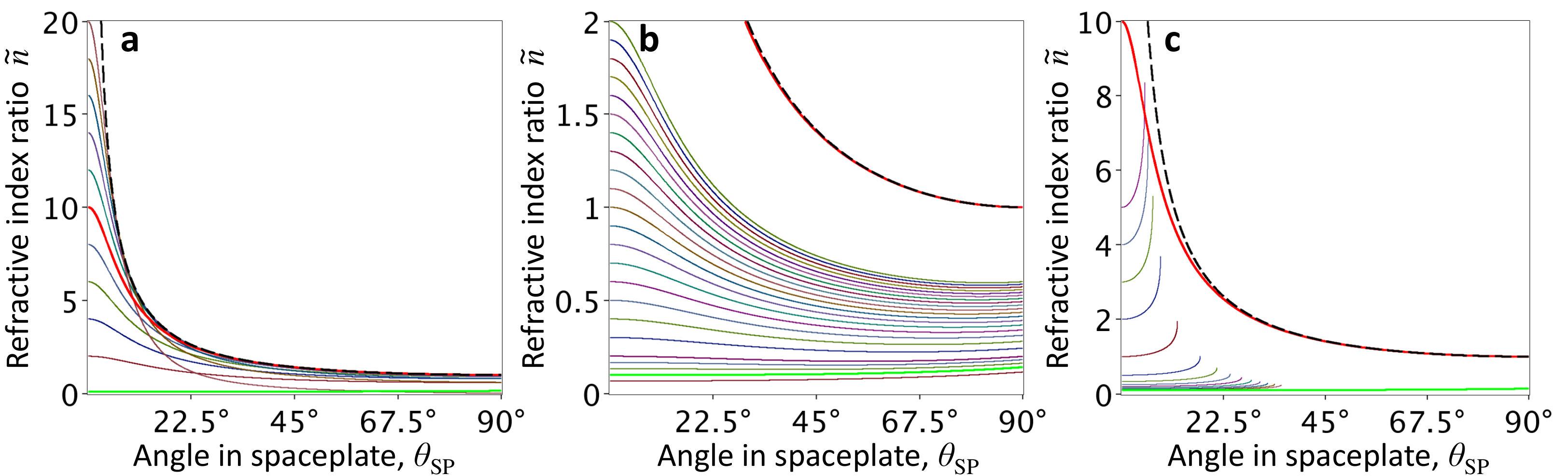}
\caption{\textbf{~|~General angle-dependent refractive index solutions for a spaceplate.} Plotted is the index ratio, $\tilde{n}\equiv n(\theta_{\mathrm{SP}})/n_{\mathrm{BG}}$. In the three panels, all the solutions are for $\mathscr{R}=10$, the black dashed line is the bound on all solutions (Eq.~\eqref{EQ:n_bound}), the red line corresponds to a uniaxial spaceplate ($C=0$, Eq.~\eqref{EQ:uniaxial_spaceplate}), and the green line corresponds to the solution approximated by the low-index spaceplate ($C=1/\mathscr{R}-\mathscr{R}$).  \textbf{a,~}Positive root solutions for $\mathscr{-R}\leq C\leq \mathscr{R}$, \emph{i.e.,} $C=\mathscr{R}(j/5-1)$ for $j=0$ to 20.  \textbf{b,~}Positive root solutions for  $C=j/\mathscr{R}-\mathscr{R}$ with $j=1$ to 10 and $3j=2$ to 6. \textbf{c,~}Negative root solutions for $C=\mathscr{R}+j$ with $j=1$ to 5 and $1/j=1$ to 10. For every positive-valued refractive index solution $\tilde{n}_+$, there is a mirror negative-valued solution, $\tilde{n}_-=-\tilde{n}_+$. }
\label{Fig:SI_gensols}
\end{figure*}

\subsection{Discussion of specific solutions}

\label{SEC:other_solutions} An infinite family of solutions are parametrized by Eq.~\eqref{EQ:SI_n_theta}. First, there are two branches to
the solution corresponding to the positive and negative roots and,
second, $C$ is arbitrary. In the following subsections, we describe
some of these solutions.

\subsubsection{Uniaxial spaceplate}
We start with the positive-root solution. We first consider the $C=0$ solution. The refractive
index ratio described by this solution is
\begin{eqnarray}
\tilde{n}\left(\theta_{\mathrm{SP}}\right) & = & \frac{\pm\sqrt{\mathscr{R}^{2}(1+\mathscr{R}^{2}\tan^{2}\theta_{\mathrm{SP}})}}{(1+\mathscr{R}^{2}\tan^{2}\theta_{\mathrm{SP}})\cos\theta_{\mathrm{SP}}}\nonumber\\
 & = & \frac{\pm\mathscr{R}}{\cos\theta_{\mathrm{SP}}\sqrt{1+\mathscr{R}^{2}\tan^{2}\theta_{\mathrm{SP}}}}.\label{EQ:n_tilde_birefringent}
\end{eqnarray}
Taking the positive root and using $\tilde{n}=n(\theta_{\mathrm{SP}})/n_{\mathrm{BG}}$, Eq.~\eqref{EQ:n_tilde_birefringent} can be re-expressed in the standard form for the extraordinary
index of a birefringent uniaxial crystal: 
\begin{align}
\frac{1}{n^{2}\left(\theta_{\mathrm{SP}}\right)} & =\frac{\cos^{2}\theta_{\mathrm{SP}}\left(1+\mathscr{R}^{2}\tan^{2}\theta_{\mathrm{SP}}\right)}{n_{\mathrm{BG}}^{2}\mathscr{R}^{2}}\\ \nonumber
 & =\frac{\cos^{2}\theta_{\mathrm{SP}}}{n_{\mathrm{BG}}^{2}\mathscr{R}^{2}}+\frac{\sin^{2}\theta_{\mathrm{SP}}}{n_{\mathrm{BG}}^{2}}\\ \nonumber
 & =\frac{\cos^{2}\theta_{\mathrm{SP}}}{n_\mathrm{o}^{2}}+\frac{\sin^{2}\theta_{\mathrm{SP}}}{n_\mathrm{e}^{2}}.
\label{EQ:uniaxial_spaceplate}
\end{align}
The last line is valid if the crystal has its extraordinary axis
along $z$, is negative uniaxial ($n_{\mathrm{o}}>n_{\mathrm{e}}$),
$n_{\mathrm{BG}}=n_{\mathrm{e}}$, and $\mathscr{R}=n_\mathrm{o}/n_\mathrm{e}$.
In this case, it will impart the ideal angle-dependent phase to mimic
$d_{\mathrm{eff}}=\mathscr{R}d$ of propagation in a medium with $n_{\mathrm{BG}}$.

Since this solution has no global phase offset ($\phi_{\mathrm{G}}=0$), not only will it act as a spaceplate for imaging, it will also replace $d_{\mathrm{eff}}$ in an interferometer. This will generally not be true; Consider, for example, an interferometer situated in a vacuum background ($n_{\mathrm{BG}}=1$). A glass plate of index $n=1.5$ and thickness $d$ can replace the `optical path length' $d_{\mathrm{eff}}=1.5d$ of interferometer arm length. Contrast this with the action of the same glass plate on the focus location of a beam propagating in vacuum. Counter to the situation in the the interferometer, refraction at the plate boundaries makes the focus shift \textit{further} along the propagation direction, as if it had passed through distance $d_{\mathrm{eff}}=({n_{\mathrm{BG}}}/{n})d=(d/1.5)$ in vacuum~\cite{Hobbs}. This is the opposite effect to what we seek, showing that the established concept of optical path length is not typically appropriate for a spaceplate. However, since the global phase offset is zero in this case, this uniaxial spaceplate will mimic propagation for the purpose of imaging \textit{and} interferometry. 

\subsubsection{Low-index spaceplate}\label{SEC:derive_lowindex}
In the family defined by
$C>(1-\mathscr{R})$,  $n(\theta_{\mathrm{SP}})$ is globally smaller
than the refractive index of the background medium, \emph{i.e.,} $n(\theta_{\mathrm{SP}})<n_{\mathrm{BG}}$.
The special case where $C=1/\mathscr{R}-\mathscr{R}$
exhibits the lowest curvature at $\theta_{\mathrm{SP}}=0$ in this family. In other words, it is the flattest
solution for small angles and, thus, has the lowest dependence on angle. Consequently, an approximation to this solution is a medium with no angular dependence at all, an isotropic medium with a refractive
index $n(0)\equiv n$. For this low-index spaceplate,
$\mathscr{R}=(n_{\mathrm{BG}}/n)$.

\subsubsection{Other solutions}

Many other solutions are possible. For each value of $C\equiv C{}_{+}$,
there is a positive root solution $\tilde{n}_{+}$, \emph{i.e.,} taking the
$+$ sign in Eq.~\eqref{EQ:SI_n_theta}. (Note, $C_{+}$ can be positive
or negative.) Paired with this positive root solution there is a mirror
negative root solution $\tilde{n}_{-}$, with $C=-C_{+}.$ It is a
mirrored about the $\tilde{n}=0$ line, in that the indices have the
same magnitude but opposite sign, $\tilde{n}_{-}=-\tilde{n}_{+}$.
A family of negative root solutions will have a refractive
index that is positive for all angles, $\tilde{n}_{-}(\theta_{\mathrm{SP}})>0$. We
plot some of these solutions in Fig.~\ref{Fig:SI_gensols}c. Critically, it is unknown
which index profiles $\tilde{n}(\theta_{\mathrm{SP}})$ are physically
allowable by Maxwell's equations.

\subsubsection{General properties of the solutions}

We now consider some limiting cases for the angle-dependent spaceplate
refractive index. We first consider a limit in which both $C$ and
$\mathscr{R}$ become large. That is, we take $C'=tC$ and $\mathscr{R'}=t\mathscr{R}$
and take the large $t$ limit:
\begin{align}
\lim_{t\to\infty}\tilde{n}\left(\theta_{\mathrm{SP}}\right)=\frac{\pm\sqrt{1-\frac{C'^{2}}{\mathscr{R'}^{2}}}}{\left|\sin\theta_{\mathrm{SP}}\right|}.
\end{align}
Consequently, in this limit, all the solutions have the same simple
angular dependence up to an overall scaling factor. 

Moreover, all the solutions, regardless of $C$ and $\mathscr{R}$, will fall between these two curves:
\begin{align}
\lim_{\mathscr{R}\to\infty}\tilde{n}\left(\theta_{\mathrm{SP}}\right)=\frac{\pm1}{\left|\sin\theta_{\mathrm{SP}}\right|},
\label{EQ:n_bound}
\end{align}
indicated by the black dashed curves in Fig.~\ref{Fig:SI_gensols}. The compression factor and $C$ parameter can be expressed in terms
of the value of $\tilde{n}$ at two angles. First, at normal incidence,
\begin{align}
\tilde{n}\left(\theta_{\mathrm{SP}}=0^\circ\right)=C\pm\mathscr{R}.
\end{align}
Second, while not all solutions give a real-valued refractive index ratio out to $\theta_{\mathrm{SP}}=90^{\circ}$, for those that do,
\begin{align}
\tilde{n}\left(90^{\circ}\right)=\pm\sqrt{1-\frac{C^{2}}{\mathscr{R}^{2}}}.
\end{align}
Consequently, in terms of these two boundaries values of $\tilde{n}$,
the compression factor is given by,
\begin{align}
\mathscr{R}=\pm\tilde{n}\left(0^\circ\right)\left(1\pm\sqrt{1-\frac{1}{\left(\tilde{n}\left(90^{\circ}\right)\right)^{2}}}\right).
\end{align}

\section{Low-index spaceplate measurements}\label{SEC:low-index_spaceplate}
Section~\ref{SEC:derive_lowindex} introduced the solution to Eq.~\ref{EQ:n_vs_theta} for which $C=(1/\mathscr{R}-\mathscr{R})$. This solution describes a spaceplate index that is lower than that of the background medium for all angles, $n(\theta_{\mathrm{SP}})<n_{\mathrm{BG}}$. It is also the flattest of any solution near $\theta_\mathrm{SP}=0$ and, thus, corresponds to an approximately isotropic medium $n(\theta_{\mathrm{SP}})\equiv n_{\mathrm{LI}}$, particularly for small incident angles. For this low-index spaceplate, $\mathscr{R}=(n_{\mathrm{BG}}/n_{\mathrm{LI}})$.

\begin{figure}[H]
    \centering
 \includegraphics[width=0.7\linewidth]{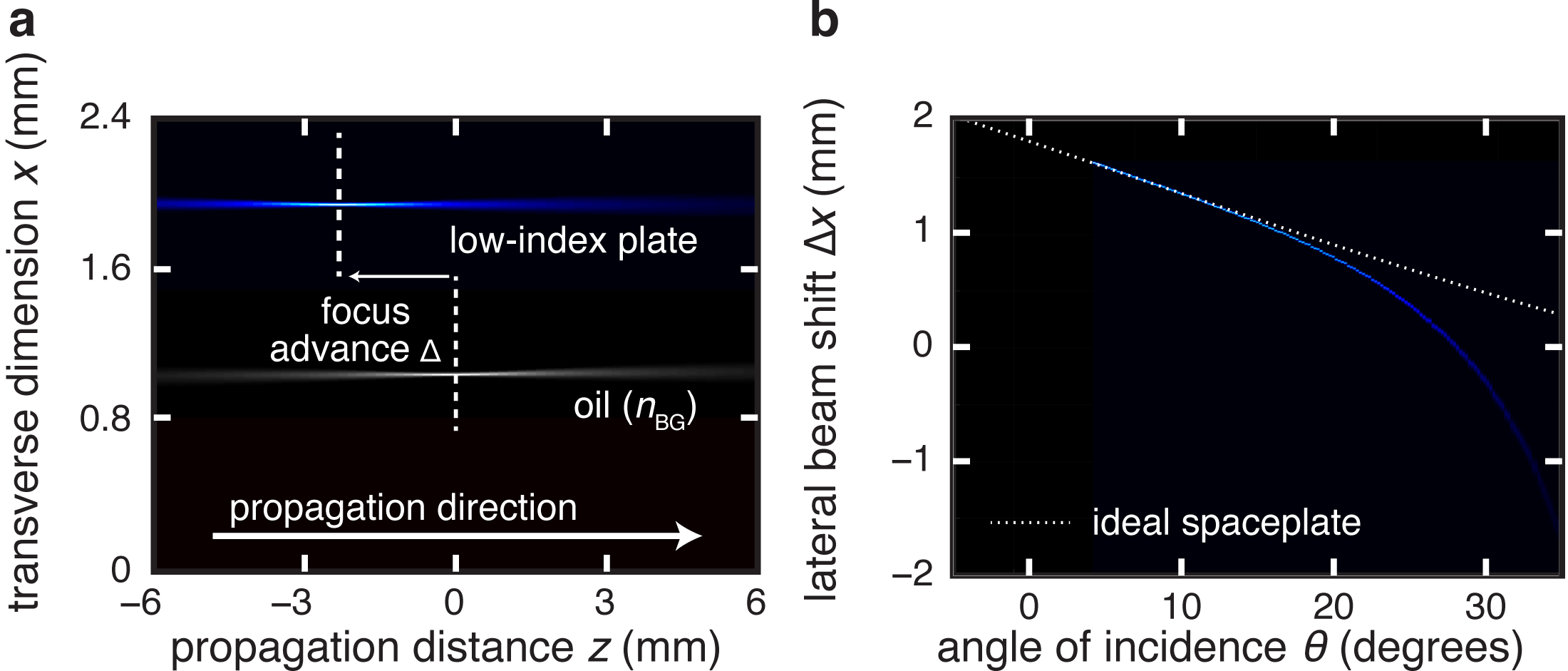}
    \caption{\textbf{~|~Space compression with a low-index spaceplate.} 
For all plots, the false-colour along the plot-vertical gives the transverse intensity distribution along $x$ at each $z$ distance on the horizontal plot axis, with paler colour corresponding to higher intensity.
\textbf{a,~}Focal shift, $\Delta=d-d_{\mathrm{eff}}$. Bottom data: Oil (grey). A converging beam comes to focus in oil at $z=0$. Top data: Low-index spaceplate (blue). Propagation through a plate of air advances the focus position along $z$  by $\Delta=-2.3$~mm. The corresponding $y$ intensity distributions are shown in Sec.~\ref{SEC:2D_spaceplate} in the Supplementary Information, demonstrating a fully two-dimensional advance.  \textbf{b,~}The walk-off of a beam incident at an angle $\theta$. The  dashed line give the lateral beam shift for an ideal spaceplate (\emph{i.e.,} $\Delta x=-(\mathscr{R}-1)d\sin\theta$ ) with the same thickness $d$ and compression factor $\mathscr{R}$ as the spaceplate in \textbf{(a)}. Above an incident angle of $\theta=15^{\circ}$, the low-index spaceplate starts to exhibit noticeable aberrations, deviating from the dotted line, due to the onset of total internal reflection and the failure of the small-angle approximation.}
    \label{Fig:low-index}
\end{figure}

If the background medium is vacuum, then $n_{\mathrm{LI}}$ must be less than one, a seemingly unusual property. Nonetheless, there exist both natural and metamaterials (\emph{e.g.,} epsilon-near-zero (ENZ) materials) from which such a spaceplate can be made\cite{Engheta2013,Vulis2018}. Current low-index ($n<1$) materials are prohibitively lossy~\cite{Engheta2013,Vulis2018}, so instead of vacuum, we select a background medium with a higher refractive index, linseed oil (which has $n_{\mathrm{BG}}=1.48$), and use air as the low-index medium. The low-index plate is a $d=4.4$-mm-long cylinder containing air and faced with glass coverslips. With air as the low index medium ($n_{\mathrm{LI}}=1$), the resulting compression factor is $\mathscr{R}=1.48$. We perform the same beam focus advance measurement as for the calcite spaceplate in the main text (as described under Beam Measurements in Sec.~\ref{SEC:Methods}). The measured shift $\Delta=-2.3$~mm agrees well with the predicted shift of $\Delta=(1-\mathscr{R})d=-2.1$~mm. 

We  next  experimentally  investigate  the  transverse  displacement  of  a  beam  incident  on  the low-index spaceplate by rotating it with respect to the incoming beam.  We see that for larger angles, the displacement induced by the low-index plate deviates from that of an ideal spaceplate (see \SI{SEC:Shift_model}). In an imaging system, this  discrepancy would act to introduce optical aberrations. This deviation is due to the failure of the small-angle approximation, most dramatically near the onset of total internal reflection at $\theta_{\mathrm{crit}}=42.5^{\circ}$. 

Aside from this aberration, total internal reflection imposes a severe limitation on the low-index spaceplate; as its refractive index $n_{\mathrm{LI}}$ decreases, its acceptance angle decreases as $\theta_{\mathrm{crit}}=\arcsin{(n_{\mathrm{LI}}/n_{\mathrm{BG}})}=\arcsin{(1/\mathscr{R})}$. Consequently, for the low-index spaceplate, the greater the compression factor $\mathscr{R}$ is, the smaller the numerical aperture (NA, \emph{i.e.,} $\theta_\mathrm{crit}$) will be. 

This measurement demonstrates a polarization-independent spaceplate effect for a slightly larger compression ($\mathscr{R}=1.48$). More importantly, it further validates our theory and Eqs.~\ref{EQ:Spaceplate}~--~\ref{EQ:n_vs_theta} in the main text.

\section{Polarization measurements}\label{SEC:polarization_calcite}
\begin{figure}[H]
    \centering
 \includegraphics[width=0.7\linewidth]{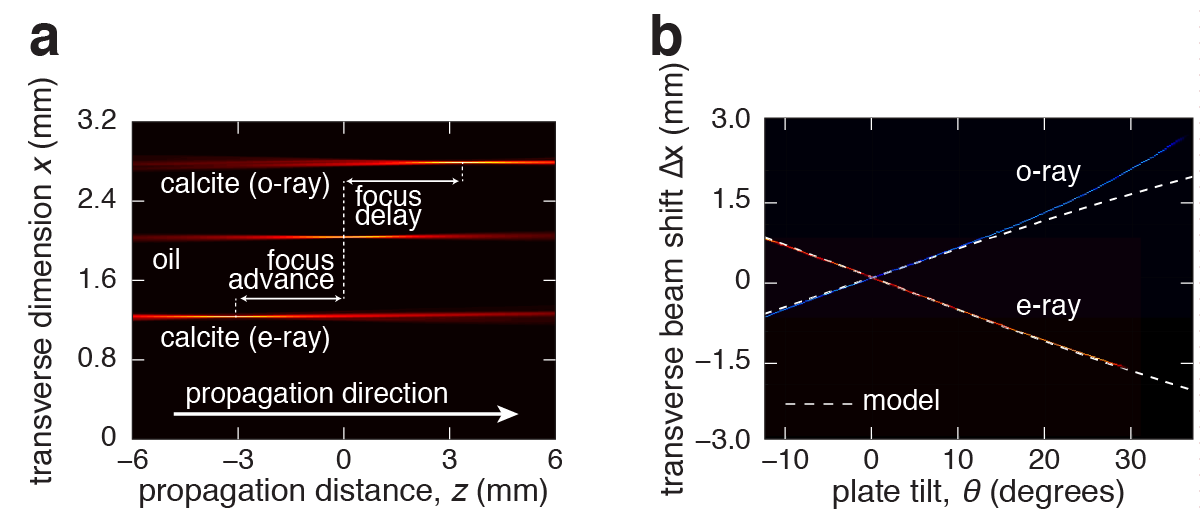}
    \caption{\textbf{~|~Polarized beam measurements with the uniaxial spaceplate.} With an extraordinarily (e-ray) polarized beam, we repeated the calcite measurements reported in Fig.~\ref{Fig:beams_measurement} (see the main paper for details). For comparison, we also include the results for an ordinarily (o-ray) polarized beam. \textbf{a,~}Focal shift $\Delta$ along $z$. Relative to the focus location in the absence of a spaceplate (oil), the o-ray has its focus shifted further from the spaceplate, \emph{i.e.,} delayed. The spaceplate effect is evident in the e-ray, which has its focus advanced towards the spaceplate.  \textbf{b,~}The lateral shift $\Delta x$ of a beam due to tilting the uniaxial spaceplate by $\theta$ relative to the $z$-axis. The o-ray shifts in opposite manner to what is required by a spaceplate, in contrast to the e-ray. The dashed lines are $\Delta x$ for an ideal spaceplate with (e-ray) compression factor $\mathscr{R}=n_\mathrm{o}/n_\mathrm{e}=1.117$ matching calcite and (o-ray) $\mathscr{R}=n_\mathrm{BG}/n_\mathrm{o}=0.895$. Notice that the lateral shift of the o-ray deviates from the approximately straight theory line whereas the e-ray does not. This deviation is evidence of the aberrations that ordinarily polarized light will encounter. In contrast, the uniaxial crystal operates ideally for all angles.}
    \label{Fig:oray}
\end{figure}

In this section, we present measurements on polarized beams with the uniaxial spaceplates. Using the polarization control detailed in Section~\ref{SEC:Methods}, we repeat the beam measurements with an ordinarily polarized beam. While for extraordinarily polarized light the uniaxial crystal acts as a spaceplate, for ordinarily polarized light it acts a homogeneous isotropic medium with refractive index $n_\mathrm{o}>n_\mathrm{BG}$. Consequently, it acts in the opposite manner to a spaceplate; this must be compensated with more propagation distance in a given setup. We measured the uniaxial crystal's effect on the focal shift of an ordinarily polarized beam and, also, on the beam's lateral shift when the spaceplate is tilted. The experimental results are shown in Fig.~\ref{Fig:oray}. The focal shift observed for the ordinarily polarized beam was  $\Delta=3.2$~mm away from the crystal which agrees well with the theoretical shift $\Delta=-(\mathscr{R}-1)d=3.126$~mm for $\mathscr{R}=n_\mathrm{BG}/n_\mathrm{o}=0.895$.

\subsection{Evidence for a fully two-dimensional spaceplate effect}\label{SEC:2D_spaceplate}
All three spaceplate types that we have introduced function in both transverse directions, $x$ and $y$. That is, the spaceplate advances the propagation of the full two-dimensional spatial distribution of the light-field.  To demonstrate this, in this section, we present the same measurement as done in Fig.~\ref{Fig:beams_measurement}a but now projecting the beam along $y$. Consequently, the plot-vertical gives the intensity distribution along $y$.

\begin{figure}[H]
\centering \includegraphics[width=0.4\textwidth]{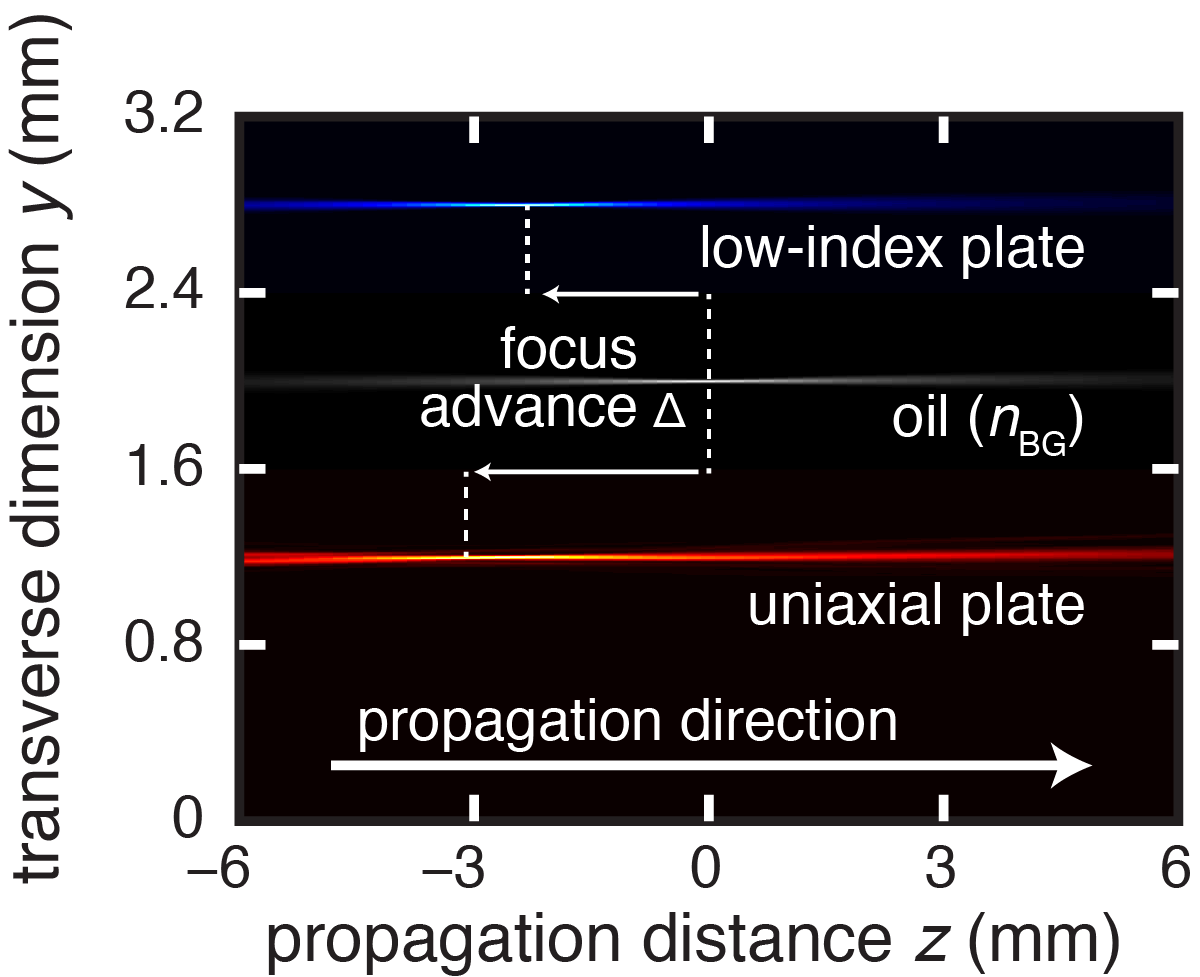}
\caption{~|~\textbf{Experimental demonstration of space compression.} 
The colour along the plot-vertical gives the transverse intensity distribution along $y$ at each point along the horizontal plot axis for all plots.
Focal shift, $\Delta=d-d_{\mathrm{eff}}$. Middle data: Oil (grey). A converging beam comes to focus in oil at $z=0$. Top data: Low-index spaceplate (blue). Propagation through a plate of air advances the focus position along $z$ by $\Delta=-2.3$~mm. Bottom data: Uniaxial spaceplate (red). Propagation of an extraordinary polarized beam through a calcite crystal with its fast axis along $z$ advances the focus position by $\Delta=-3.4$~mm.}
\label{Fig:beams_measurement_y}
\end{figure}

\section{Movies} \label{SEC:visualizations}
Below are the captions for the movies.

\begin{itemize}

\item Propagation of a beam in both oil and the air plate (Movie 1) and in both oil and calcite (Movie 2) as a function of beam propagation distance in the oil~$z$. The green laser beam looks red due to post-processing. These videos were integrated to produce Fig.~\ref{Fig:beams_measurement}.

\item Propagation of a white-light-illuminated image in both glycerol and calcite (Movie 3) as a function of beam propagation distance in the glycerol~$z$.
\end{itemize}

\section{Implications of causality}\label{SEC:Causality}

Causality does not appear to limit the operation of a possible spaceplates with regards to the maximum achievable compression factors $\mathscr{R}$. When extrapolating from the demonstrations above, it is evident that a spaceplate can be designed to exhibit an arbitrarily large $\mathscr{R}$ for a single operating frequency by employing a medium with a sufficiently low refractive index. Instead, this system reveals an explicit trade-off between $\mathscr{R}$ and the numerical aperture of the spaceplate ($\mathrm{NA}= (1/\mathscr{R}$)). However, whether this trade-off is inherent to the spaceplate concept or is a consequence of the low-index spaceplate implementation remains an open question. A broadband spaceplate based on an index less than unity would exhibit a violation of microscopic causality, an energy velocity at a point inside the medium that is higher than $c$. Thus, the bandwidth of a low-index spaceplate will also be constrained.

In contrast, a uniaxial spaceplate designed to work in vacuum ($n_\mathrm{e}=1$, $n_\mathrm{o}=\mathscr{R}$) will not violate  microscopic causality; its operation depends on the ratio of its two refractive indices, and these two quantities are not causally related. To this point, the group index in a non-dispersive uniaxial material is always bounded by its birefringent indices, $n_\mathrm{e}$ and $n_\mathrm{o}$, and, thus, the group velocity will always be less than $c$. Indeed, our experimental demonstration supports the prospect of full broadband operation of a spaceplate. Microscopic causality also does not exclude the possibility of spaceplate operation in free space that is based on anisotropic materials.

This conclusion may seem to conflict with intuition derived from experience with other transformation optics-based devices~\cite{Monticone2013a}, such as an invisibility cloak. A functioning cloak necessitates light travelling within a medium to keep pace with light propagating in the surrounding background. This restriction on macroscopic causality may be relaxed by immersing the device within a high-index medium since then light propagating in the surrounding medium will be slowed. By contrast, a spaceplate encompasses the entire imaging beam, and it has no physical boundary in the direction transverse to the direction of propagation to impose such a strict condition. Therefore, there is no penalty for a beam propagating through a spaceplate to be delayed with respect to another beam propagating in free space alongside it. Consequently, the overall time delay for the imaging light to pass through the spaceplate is unconstrained, becoming a free parameter in spaceplate design. For this reason, we do not foresee any intrinsic issues with limited operation bandwidth in the spaceplate platform, despite evidence to the contrary in other systems. 

We suspect that rigorous bandwidth limits could be derived for specific implementations of spaceplates using methods that have already been applied to metasurfaces (\emph{e.g.,} those described in Ref.~\cite{Presutti2020a}). These investigations are of interest for future research.

If causality was ultimately revealed to pose a restriction on broadband operation, a spaceplate operating at a single frequency would still be beneficial towards many applications. For instance, there is a great deal of interest in wavefront shaping applications that make use of the transverse degree of freedom of light, such as spatial multiplexing in telecommunications~\cite{Bozinovic2013}, or high-dimensional quantum cryptography~\cite{Sit2017}. Applications of these type notably operate at a single frequency and could usually benefit from being compressed by a spaceplate. Additionally, as is argued in the main text, many imaging applications only use three colours centered around red, green, and blue frequencies to enable full-colour imaging. A spaceplate that can perform simultaneous compressions for three discrete frequencies could enable full-color imaging without the need of being broadband. The question of broadband operation or causality can, fortunately, be circumvented for these applications.

\section{Fabricated spaceplates}\label{SEC:physical_spaceplates}
\begin{figure}[H]
    \centering \includegraphics[width=89mm]{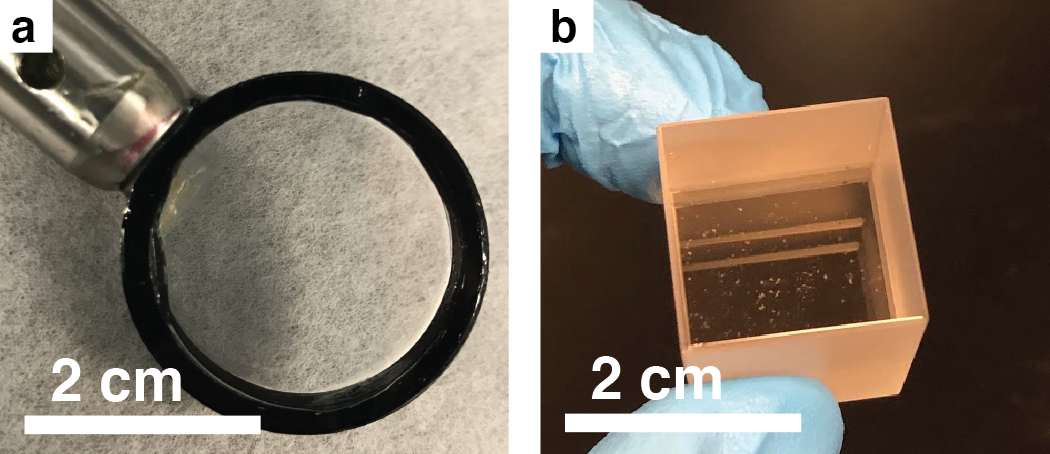}
    \caption{\textbf{~|~Fabricated spaceplates. (A)~}Low-index spaceplate. \textbf{b,~}Uniaxial spaceplate. The entrance and exit surfaces exhibit visible roughness.}
    \label{Fig:spaceplates}
\end{figure}

The low-index spaceplate was fabricated by attaching microscope cover slips to both ends of a lens tube using epoxy (Fig.~\ref{Fig:spaceplates}a). The uniaxial spaceplate was fabricated by a commercial vendor, which informed us that, in calcite, surfaces normal to the extraordinary axis are difficult to polish well (Fig.~\ref{Fig:spaceplates}b). Consequently, some roughness can be observed on the entrance and exit surfaces, resulting in scattering, particularly in the beam measurements.

\section{Transverse and lateral beam shifts due to a spaceplate}

\label{SEC:Shift_model}

\begin{figure}[H]
\centering \includegraphics[width=70mm]{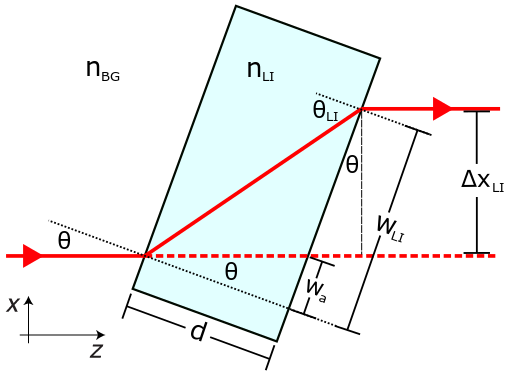}
\caption{\textbf{~|~Geometry for the derivation of lateral beam shift $\Delta x_{\mathrm{LI}}$ in the case of a low-index plate.}}
\label{Fig:low_index_derivation} 
\end{figure}

As depicted in Fig.~\ref{Fig:schematic}a in the paper, one expects a spaceplate to shift a beam of light when incident at an angle~$\theta.$ In this section, we derive the transverse shift $w$ and lateral shift $\Delta x$. We first derive the shift due to propagating through a slab of medium
$n_{\mathrm{BG}}$ of thickness $d_{\mathrm{eff}}$, which defines
the effect of an ideal spaceplate. We then derive these shifts for
the low-index spaceplate and uniaxial spaceplate.

\subsection{Beam shifts due to an ideal spaceplate}

Consider a beam traveling in the $x,z$ plane at an angle $\theta$
to the $z$-axis. The beam traverses a region (\emph{i.e.,} a slab) of a
medium at length $z=d_{\mathrm{eff}}$, entering at point $(x_\mathrm{in},z=0)$
and exiting at point ($x_\mathrm{out},z=d_{\mathrm{eff}}$). Upon exiting the region the beam is parallel to the beam that entered the region. That is, its angle is unchanged. 

By simple geometry the beam shifts in the $x$-direction by an amount,
\begin{equation}
x_\mathrm{out}-x_\mathrm{in}=-d_{\mathrm{eff}}\tan{\theta}\equiv w,\label{EQ:ideal_transverse}
\end{equation}
which we name the `transverse' shift. The negative sign is due to a convention that will be explained in the next paragraph. For an ideal spaceplate, the
same beam will exit translated along the plate's interface by an equal
amount $w$ of transverse shift as created by the slab of medium described
above. Another way of looking at the transverse shift is as a displacement from a line that both goes through point $(x_\mathrm{in},z=0)$
and is normal to the plate interfaces. In contrast to the scenario with the medium, this shift now occurs at a distance $z=d$, the plate thickness.

To test this effect, rather than tilt the incoming beam by $\theta$,
it is experimentally simpler to tilt the studied spaceplate by the
same angle. In this case, the quantity that can be measured most directly
is the `lateral' shift $\Delta x$ of the beam. This is the shift
of the beam along an axis normal to the beam's propagation direction. The sign convention mentioned in the last paragraph sets $\Delta x$ to be positive if the lateral shift is below the path of the incident beam, as depicted in Fig.~\ref{Fig:low_index_derivation}. In particular, $\Delta x$ would be positive for a beam passing through a tilted regular glass plate in a vacuum background medium.   

To relate $\Delta x$ to $w$, one must account for the tilt of the normal
of the plate interface. This alone would result in apparent transverse
shift of $w_{a}=d\tan{\theta}$, which must be added to the actual transverse
shift $w.$ With this established, geometry gives the lateral shift as,
\begin{align}
\Delta x= & \left(w+w_{a}\right)\cos\theta\nonumber \\
= & \left(-d_{\mathrm{eff}}\tan{\theta}+d\tan{\theta}\right)\cos\theta\nonumber \\
= & -d\left(\mathscr{R}-1\right)\sin\theta,\label{EQ:ideal_lateral}
\end{align}
where we have used $d_{\mathrm{eff}}=\mathscr{R}d$.

\subsection{Beam shifts due to a low-index spaceplate}

We now consider the lateral shift created by a plate made of an isotropic
and homogeneous medium of thickness $d$. If the medium's refractive
index $n_{\mathrm{LI}}<n_{\mathrm{BG}}$, then the plate is a low-index
spaceplate. To start, however, we leave $n_{\mathrm{LI}}$ unrestricted. At the entrance interface, incoming light refracts to angle
$\theta_{\mathrm{LI}}$ inside the plate according to Snell's law.
In the following, we use Snell's law to derive analogous relations
for $\cos$ and $\tan$:
\begin{align}
\sin{\theta_{\mathrm{LI}}} & =\frac{n_{\mathrm{\mathrm{BG}}}}{n_{\mathrm{\mathrm{LI}}}}\sin{\theta},\nonumber \\
\cos{\theta_{\mathrm{LI}}} & =\sqrt{1-\sin^{2}\theta_{\mathrm{LI}}}=\sqrt{1-\left(\frac{n_{\mathrm{\mathrm{BG}}}}{n_{\mathrm{\mathrm{LI}}}}\right)^{2}\sin^{2}\theta},\nonumber \\
\tan{\theta_{\mathrm{LI}}} & =\frac{\sin{\theta_{\mathrm{LI}}}}{\cos{\theta_{\mathrm{LI}}}}=\frac{\frac{n_{\mathrm{\mathrm{BG}}}}{n_{\mathrm{\mathrm{LI}}}}\sin{\theta}}{\sqrt{1-\left(\frac{n_{\mathrm{\mathrm{BG}}}}{n_{\mathrm{\mathrm{LI}}}}\right)^{2}\sin^{2}\theta}}=\frac{\sin{\theta}}{\sqrt{\left(\frac{n_{\mathrm{\mathrm{LI}}}}{n_{\mathrm{\mathrm{BG}}}}\right)^{2}-\sin^{2}\theta}}.\label{EQ:tanDeriv}
\end{align}
As in Eq.~\eqref{EQ:ideal_transverse} from the last section, simple geometry shows that the transverse shift of the beam is,
\begin{align}
w_{\mathrm{LI}}=-d\tan{\theta_{\mathrm{LI}}}.
\end{align}
Likewise, the lateral shift is, 
\begin{align}
\Delta x_{\mathrm{LI}}= & \left(w_{\mathrm{LI}}+w_{a}\right)\cos\theta\nonumber \\
= & \left(-d\tan{\theta_{\mathrm{LI}}}+d\tan{\theta}\right)\cos\theta.\label{EQ:EQ:delta_x}
\end{align}
Substituting Eq.~\eqref{EQ:tanDeriv} into Eq.~\eqref{EQ:EQ:delta_x}
yields an expression for the lateral shift that is dependent only
on~$\theta$,
\begin{align}
\Delta x_{\mathrm{LI}}=-d\sin{\theta}\left(\frac{\cos{\theta}}{\sqrt{\left(\frac{n_{\mathrm{\mathrm{LI}}}}{n_{\mathrm{\mathrm{BG}}}}\right)^{2}-\sin^{2}\theta}}-1\right).
\end{align}
This shows that the lateral shift created by a low-index spaceplate is not equal
to the ideal lateral shift given in Eq.~\eqref{EQ:ideal_lateral}. However, using the small angle approximation ($\theta\ll1$) we find that
\begin{align}
\Delta x_{\mathrm{LI}} & \approx -d\sin{\theta}\left(\frac{1}{\sqrt{\left(\frac{n_{\mathrm{\mathrm{LI}}}}{n_{\mathrm{\mathrm{BG}}}}\right)^{2}-0}}-1\right) +O\left(\theta^3\right)\nonumber \\ 
 & =-d\sin{\theta}\left(\left|\frac{n_{\mathrm{\mathrm{BG}}}}{n_{\mathrm{\mathrm{LI}}}}\right|-1\right) \nonumber\\ 
 & =-d\left(\mathscr{R}-1\right)\sin\theta \nonumber\\ 
 & =\Delta x, \label{EQ:low_index_lateral}
\end{align}
where we have identified the low-index compression factor, $\mathscr{R}=|n_{\mathrm{\mathrm{BG}}}/n_{\mathrm{\mathrm{LI}}}|$.
For large angles, the low-index spaceplate will produce an incorrect
shift, effectively introducing aberrations. Moreover, all angles above
the critical angle, $\theta_{c}=\arcsin(n_{\mathrm{\mathrm{LI}}}/n_{\mathrm{BG}}),$
are perfectly reflected. However, for small angles a low-index plate
acts as an ideal spaceplate, \emph{i.e.,} $\Delta x_{\mathrm{LI}}=\Delta x$.

While we have derived the lateral shift $\Delta x_{\mathrm{LI}}$
in the context of a low-index spaceplate, it is actually valid for
any value of $n_{\mathrm{LI}}$, including a negative refractive index.
Moreover, we use it to describe the action of the uniaxial spaceplate
on ordinarily polarized light, which effectively experiences a homogeneous
isotropic medium with index $n_\mathrm{o}$. Lastly, the shift $\Delta x_{\mathrm{LI}}$ also shows
that a simple glass plate in air will introduce imaging aberrations
for large angles.

\subsection{Beam shifts due to a uniaxial spaceplate}

We will now derive the lateral shift created by a uniaxial spaceplate.
The uniaxial crystal is characterized by two indices, the extraordinary
and ordinary refractive indices, $n_\mathrm{e}$ and $n_\mathrm{o}$, respectively.
To derive the shift, one must account for the anomalous refraction
and consequent beam walk-off that generally occurs with a birefringent
medium. In particular,  for extraordinarily polarized light in a uniaxial crystal (U), the Poynting vector
$\mathbf{S}^{(\mathrm{U})}$, which describes the energy flow, can point
in a direction different from the wavevector $\mathbf{k}^{(\mathrm{U})}$. In contrast, an ordinarily polarized beam will refract normally (\emph{i.e.,} the Poynting and wavevectors are parallel), and, hence, will shift according to Eq.~\eqref{EQ:low_index_lateral} but with $n_{\mathrm{\mathrm{LI}}}$ replaced with $n_\mathrm{o}.$ In the following, we use the results from Ref.~[\cite{Entezar_2019}], which carefully analyzed the Poynting vector angle at the interface between a homogeneous medium and a uniaxial crystal with its extraordinary axis (\emph{i.e.,} optic axis) at an arbitrary angle. Since the Poynting vector is the direction an extraordinary beam travels along in the crystal, one can use it to follow a geometric derivation similar to that used in Eq.~\eqref{EQ:ideal_lateral}.

As discussed earlier in the SI, the transverse component of the wavevector
is conserved across the interface, 
\begin{equation}
k_{\perp}^{(\mathrm{U)}}=k_{\perp}^{(\mathrm{BG)}}\equiv k_{\perp}=k_{0}n_{\mathrm{\mathrm{BG}}}\sin\theta.\label{EQ:k_transverse}
\end{equation}
We apply the results from Ref.~[\cite{Entezar_2019}] to the case where the extraordinary
axis is normal to the plate interface. In this case, the $z$-component
of the wavevector in the crystal can be expressed as:
\begin{equation}
k_{z}^{\mathrm{(U)}}=n_\mathrm{o}k_{0}\sqrt{1-\left(\frac{n_{\mathrm{\mathrm{BG}}}}{n_\mathrm{e}}\right)^{2}\sin^{2}\theta}.\label{EQ:k_z}
\end{equation}
If one considers the case where $n_{\mathrm{BG}}=n_{\mathrm{e}}$,
then $k_{z}^{\mathrm{(U)}}=n_\mathrm{o}k_{0}\cos\theta$, which directly shows why a uniaxial crystal acts as a spaceplate: it produces an angle-dependent phase that is magnified by a factor $n_\mathrm{o}/n_\mathrm{e}$
relative to propagation through a medium with $n_{\mathrm{BG}}=n_{\mathrm{e}}.$
For now, however, we leave $n_\mathrm{e}$ general.

Whereas, after refraction at the interface, the wavevector angle in the
crystal will be given by $\tan\theta_{\mathrm{U},k}=k_{\perp}^{\mathrm{(U)}}/k_{z}^{\mathrm{(U)}}$,
Ref.~[\cite{Entezar_2019}] showed that the Poynting vector of an extraordinarily polarized
plane-wave will be along a potentially different angle,
\begin{align*}
\tan\theta_{\mathrm{U},S} & =\frac{n_\mathrm{o}^{2}k_{\perp}^{\mathrm{(U)}}}{n_\mathrm{e}^{2}k_{z}^{\mathrm{(U)}}}\\
 & =\frac{n_\mathrm{o}^{2}k_{0}n_{\mathrm{\mathrm{BG}}}\sin\theta}{n_\mathrm{e}^{2}n_\mathrm{o}k_{0}\sqrt{1-\left(\frac{n_{\mathrm{\mathrm{BG}}}}{n_\mathrm{e}}\right)^{2}\sin^{2}\theta}},
\end{align*}
where we have used the expressions for the two wavevector components,
Eqs.~\eqref{EQ:k_transverse} and \eqref{EQ:k_z}.

From here on, we restrict ourselves to the case in which $n_{\mathrm{BG}}=n_{\mathrm{e}}$.
In this case, the Poynting vector angle reduces to 
\begin{equation}
\tan\theta_{\mathrm{U},S}=\mathscr{R}\tan\theta,\label{EQ:uniaxial_angle}
\end{equation}
where we have identified the compression factor as $\mathscr{R}=n_\mathrm{o}/n_\mathrm{e}$,
as expected from our phase analysis earlier. Using Eq.~\eqref{EQ:uniaxial_angle}, the transverse shift for an extraordinarily polarized beam will be 
\begin{equation}
w_{\mathrm{U}}=-d\mathscr{R}\tan\theta \label{EQ:uniaxial_transverse},
\end{equation}
which can be used to find the associated lateral shift:
\begin{align}
\Delta x_\mathrm{U}= & \left(w_{\mathrm{U}}+w_{a}\right)\cos\theta\nonumber \\
= & \left(-d\tan\theta_{\mathrm{U},S}+d\tan{\theta}\right)\cos\theta\label{EQ:uniaxial_lateral}\\
= & -d\left(\mathscr{R}-1\right)\sin\theta\nonumber \\
= & \Delta x.\nonumber 
\end{align}
This shows that for all angles, the lateral shift of an extraordinarily
polarized beam $\Delta x_\mathrm{U}$ will be identical to the lateral shift
of an ideal spaceplate $\Delta x$. In this sense, a uniaxial crystal
acts as a perfect spaceplate for the purposes of imaging, replacing medium $n_\mathrm{BG}$, while introducing
no aberrations.

\end{document}